\documentclass[twocolumn]{aastex631}
\usepackage{CJK} 

\usepackage{amsmath}
\usepackage{amssymb}
\usepackage{epsfig}
\usepackage{apjfonts}
\usepackage{natbib}
\usepackage{hyperref}
\usepackage{epstopdf}
\usepackage{verbatim}
\usepackage{enumitem}
\usepackage{color}
\setlist[enumerate]{itemsep=-1mm}
\usepackage{soul, xcolor}
\setstcolor{blue}

\usepackage{wasysym}

\usepackage{textcase}

\usepackage{mathtools}

\usepackage{bigints}

\newcommand{\ulsst}{u_{\rm LSST}}
\newcommand{\glsst}{g_{\rm LSST}}
\newcommand{\rlsst}{r_{\rm LSST}}
\newcommand{\ilsst}{i_{\rm LSST}}
\newcommand{\zlsst}{z_{\rm LSST}}
\newcommand{\ylsst}{y_{\rm LSST}}
\newcommand{\yvhs}{Y_{\rm VHS}}
\newcommand{\jvhs}{J_{\rm VHS}}
\newcommand{\hvhs}{H_{\rm VHS}}
\newcommand{\kvhs}{K_{S,\rm VHS}}
\newcommand{\wa}{W1_{\rm CatWISE}}
\newcommand{\wb}{W2_{\rm CatWISE}}
\newcommand{\gps}{g_{\rm P1}}
\newcommand{\rps}{r_{\rm P1}}
\newcommand{\ips}{i_{\rm P1}}
\newcommand{\zps}{z_{\rm P1}}
\newcommand{\yps}{y_{\rm P1}}
\newcommand{\jmass}{J_{\rm 2MASS}}
\newcommand{\hmass}{H_{\rm 2MASS}}
\newcommand{\kmass}{K_{S,\rm 2MASS}}
\newcommand{\ymko}{Y_{\rm MKO}}
\newcommand{\jmko}{J_{\rm MKO}}
\newcommand{\hmko}{H_{\rm MKO}}
\newcommand{\kmko}{K_{\rm MKO}}
\newcommand{\ieuclid}{I_{\rm E}}
\newcommand{\yeuclid}{Y_{\rm E}}
\newcommand{\jeuclid}{J_{\rm E}}
\newcommand{\heuclid}{H_{\rm E}}

\submitjournal{AJ on August 22, 2025 (Accepted: October 2, 2025)}

\begin{document}


\begin{CJK*}{UTF8}{gbsn}

\title{Ultracool dwarf Science with MachIne LEarning (USMILE). I. Scalable Tree-Based Models for Photometric Spectral Classification and New Discoveries from LSST Data Preview 1 and Euclid Quick Data Release 1\footnote{This paper is dedicated to our daughter, Allison. The model introduced here, \texttt{USMILE Avocado}, is named after Allison's very first solid food. As parents, we hope U~SMILE every day.}}

\author[0000-0002-3726-4881]{Zhoujian Zhang (张周健)}
\affiliation{Department of Physics \& Astronomy, University of Rochester, Rochester, NY 14627, USA}

\author[0009-0003-2773-5867]{Yanxia Li (李燕侠)}
\affiliation{Amazon.com, Inc., 410 Terry Ave N, Seattle, WA 98109, USA}

\begin{abstract}
We present the Ultracool~dwarf~Science~with~MachIne~LEarning (USMILE), a program developing machine-learning tools for the discovery and characterization of ultracool dwarfs. We introduce \texttt{USMILE~Avocado}, a spectral classification framework that uses broadband photometry from wide-field surveys --- Rubin~Observatory LSST~Data~Preview~1 (DP1), VISTA Hemisphere Survey (VHS), and CatWISE --- as input features. The framework comprises two gradient-boosted decision-tree models scalable to the massive data volumes of modern surveys: the \texttt{classifier}, which distinguishes ultracool dwarfs from stellar/extragalactic contaminants, and the \texttt{regressor}, which predicts spectral types. A key strength is its ability to natively handle missing photometric features, common in wide-field searches, whereas earlier machine-learning approaches required complete multi-band detections or relied on imputation, thereby excluding genuine ultracool dwarfs or introducing bias. Trained on an augmented labeled dataset of $>2$ million sources built from known ultracool dwarfs, reddened early-type stars, and quasars, the models achieve strong performance: the \texttt{classifier} attains an ROC~AUC of 0.976 and an F1~score of 0.92, while the \texttt{regressor} yields a mean-squared error of 0.88~subtypes. Applying these models, we carried out the first ultracool dwarf search with LSST~DP1, cross-matched against VHS~and~CatWISE. Crucially, Euclid Quick~Data~Release~1 provided near-infrared spectra for hundreds of candidates, enabling a rare, large-scale external spectroscopic validation. This confirmed 15~M6--L2 discoveries, verified \texttt{USMILE} performance, and clarified regimes where \texttt{USMILE} predictions are most reliable. Building on these insights, we identified 25 additional high-quality M6--L9 photometric candidates. These early discoveries demonstrate the effectiveness of scalable machine-learning methods in the data-rich era of wide-field surveys, highlighting the synergy between LSST and Euclid in expanding the ultracool dwarf census.
\end{abstract}

\section{Introduction} 
\label{sec:introduction}

Ultracool dwarf science has advanced fundamentally through wide-field digital sky surveys, which serve as the primary discovery engines for low-mass stars, brown dwarfs, and planetary-mass objects. Notable examples include the Sloan Digital Sky Survey \citep[SDSS;][]{2000AJ....120.1579Y}, Gaia \citep{2016A&A...595A...1G}, the Panoramic Survey Telescope and Rapid Response System 1 survey \citep[Pan-STARRS1;][]{2016arXiv161205560C}, Dark Energy Survey \citep[DES;][]{2018ApJS..239...18A}, Two Micron All-Sky Survey \citep[2MASS;][]{2006AJ....131.1163S}, the UKIRT Infrared Deep Sky Survey \citep[UKIDSS;][]{2007MNRAS.379.1599L}, VISTA Hemisphere Survey \citep[VHS;][]{2013Msngr.154...35M}, and several catalogs produced by the Wide-field Infrared Survey Explorer \citep[WISE;][]{2010AJ....140.1868W}, such as AllWISE \citep{2012yCat.2311....0C} and CatWISE \citep{2020ApJS..247...69E, 2021ApJS..253....8M}. These surveys provide broadband photometry spanning from the optical to the mid-infrared, along with multi-epoch astrometry, enabling the discovery and characterization of ultracool dwarfs across broad ranges of effective temperature, mass, age, metallicity, and birth environment \citep[e.g.,][]{1988Natur.336..656B, 
1991AJ....101..662B, 
1993ApJ...414..279T, 2005AJ....130.2326T, 2018ApJS..236...28T,
1995Natur.378..463N, 
1995AJ....110.1838R, 2003AJ....126.3007R, 2008AJ....136.1290R,  
1997ApJ...476..311K, 1999ApJ...519..802K, 2000AJ....120..447K, 2008ApJ...689.1295K, 2010ApJS..190..100K, 2011ApJS..197...19K, 2012ApJ...753..156K, 2014ApJ...783..122K, 2016ApJS..224...36K, 2021ApJS..253....7K, 
1997A&A...327L..25D, 
1999ApJ...522L..65B, 2002ApJ...564..421B, 2003ApJ...594..510B, 2003ApJ...586..512B, 2003ApJ...592.1186B, 2004AJ....127.2856B,
1999AJ....118.2466M, 2010A&A...517A..53M, 
1999ApJ...522L..61S,
2000AJ....119..928F, 
2000A&A...353..958S, 2004A&A...425..519S, 2009A&A...494..949S, 
2000ApJ...531L..61T, 
2000ApJ...536L..35L, 
2000AJ....120.1085G, 2001ApJ...551L.163G, 
2000Sci...290..103Z, 
2001A&A...380..590P, 2008MNRAS.383..831P, 
2002ApJ...564..466G, 
2002AJ....123.3409H, 
2002AJ....124.1190L, 2003AJ....125.1598L, 
2002AJ....123..458S, 2015ApJ...804...92S, 2016ApJ...817..112S, 2017AJ....153..196S, 2022AJ....163..242S, 
2003AJ....126.2421C, 2007AJ....133..439C, 
2003A&A...403..929K, 2004A&A...416L..17K, 2004AJ....127.3553K, 
2003ApJ...590..348L, 2003ApJ...593.1093L, 2005ApJ...618..810L, 2011ApJ...730L...9L, 2012ApJ...760..152L, 2016ApJ...827...52L, 2018AJ....156...76L, 2024ApJ...975..162L, 2004ApJ...616.1033L, 2014ApJ...786L..18L, 2023AJ....165..269L, 2024AJ....168..159L, 
2003IAUS..211..197W,
2005A&A...440.1061L, 2007MNRAS.379.1423L, 2008MNRAS.383.1385L, 2012A&A...542A.105L, 2014MNRAS.445.3908L, 
2006AJ....131.2722C, 
2007AJ....134.1162L, 
2008ApJ...676.1281M, 
2008MNRAS.390..304P, 
2008AJ....135..785W, 2011AJ....141...97W, 
2009AJ....137....1F, 
2009A&A...497..619Z, 2010MNRAS.404.1817Z, 2019MNRAS.486.1260Z, 
2010MNRAS.406.1885B, 2013MNRAS.433..457B, 
2010AJ....139.1808S, 
2011AJ....141..203A, 
2011ApJ...743...50C, 
2011ApJ...732...56G, 
2013A&A...549A.123A, 
2013MNRAS.430.1171D, 
2013ApJ...777L..20L, 
2013ApJ...777...84B, 2015ApJ...814..118B, 2017ApJ...837...95B, 2020AJ....159..257B, 
2013ApJS..205....6M, 
2013AJ....146..161M, 2015MNRAS.449.3651M, 
2013PASP..125..809T, 
2014ApJ...792..119D, 
2014MNRAS.443.2327S, 
2015MNRAS.450.2486C, 
2015ApJS..219...33G, 
2015AJ....150..182K, 2017AJ....154..112K,
2016ApJ...830..144R, 
2016A&A...589A..49S,
2018A&A...619L...8R, 
2018ApJ...858...41Z, 2020ApJ...891..171Z, 2021ApJ...916L..11Z, 2022ApJ...935...15Z,
2019MNRAS.489.5301C,
2019AJ....158..182G, 
2020ApJ...905L..14F, 
2020ApJ...892..122J, 
2020ApJ...889...74M, 2020ApJ...899..123M, 
2021AJ....162..102S, 
2023MNRAS.522.1951D, 
2023ApJ...951..139D, 
2023ApJ...959...63S}. Beyond individual discoveries, they have produced large statistical samples that constrain the luminosity and mass functions, birth rates, and spatial distributions of ultracool dwarfs across the Galaxy \citep[e.g.,][]{1999ApJ...525..440L, 2000ApJ...533..358M, 2000ApJ...540..236H, 2005ApJ...625..385A, 2008MNRAS.390..304P, 2010A&A...522A.112R, 2012ApJ...753..156K, 2021ApJS..253....7K, 2013MNRAS.433..457B, 2013MNRAS.430.1171D, 2024ApJ...967..115B}.

A transformative new era is now underway with the NSF-DOE Vera C. Rubin Observatory's Legacy Survey of Space and Time \citep[LSST;][]{2019ApJ...873..111I} and ESA's Euclid mission \citep{2025A&A...697A...1E}. LSST, conducted with the 8.4-m Simonyi Survey Telescope, will provide deep, multi-epoch imaging of the southern hemisphere in six optical bands ($\ulsst/\glsst/\rlsst/\ilsst/\zlsst/\ylsst$) over the coming decade. On June 30, 2025, the Rubin Observatory released LSST Data Preview 1 \citep[DP1;][]{2025lsst.data....3N, 2025rubn.rept...31N}, a subset of commissioning data covering seven $\sim 1$~degree$^{2}$ fields with varying depths across the filters \citep{Guy2025EarlyScience}. In parallel, Euclid is delivering wide-field optical imaging ($\ieuclid$), near-infrared imaging ($\yeuclid/\jeuclid/\heuclid$), and slitless near-infrared spectroscopy across more than one-third of the sky. Its first public release, Euclid Quick Data Release 1 \citep[Q1;][]{2025arXiv250315302E}, became available on March 19, 2025, and includes nearly 63~degree$^{2}$ spectrophotometric observations from Euclid Deep Fields South (EDF-S), North (EDF-N), and Fornax (EDF-F). Notably, one LSST DP1 field lies within the EDF-S, creating an early opportunity to explore synergies between the two surveys. Together, these data releases offer unprecedented opportunities to uncover a larger and more diverse population of ultracool dwarfs, while also providing early testbeds for methods that can effectively identify ultracool dwarfs from these data and scale to the massive data volumes of LSST and Euclid.

Identifying ultracool dwarfs in these wide-field surveys is, however, nontrivial given the immense scale of the datasets and the prevalence of contaminants. These contaminants include reddened early-type stars, extragalactic sources (e.g., quasars), spurious detections from detector artifacts, and objects with compromised photometry due to blending or poor data quality. Traditional searches have relied on simple color cuts to isolate red sources. While effective and straightforward, this method produces large candidate lists dominated by contaminants and often requires additional filtering steps, such as visual inspection of imaging data \citep[e.g.,][]{2011ApJS..197...19K}, visual checks of color-color and color-magnitude diagram positions \citep[e.g.,][]{2020ApJ...891..171Z}, proper motion analysis \citep[e.g.,][]{2023MNRAS.522.1951D}, or requirements that candidates be detected across multiple surveys \citep[e.g.,][]{2015ApJ...814..118B}. Some of these approaches are manual and cannot easily scale to modern surveys like LSST and Euclid, where data volumes reach hundreds of petabytes. Tightening the color cuts can reduce contamination, but only at the cost of increasing false negatives and excluding genuine ultracool dwarfs.

Moreover, color cuts cannot infer candidates' properties, such as spectral types, which provide critical context for prioritizing spectroscopic follow-up and deriving population-level statistics. To address this limitation, \cite{2015A&A...574A..78S} introduced the photo-type method, which estimates spectral types directly from multi-band photometry. In this approach, a single template spectral energy distribution (SED) is constructed for each subtype from M5--T8 using photometry from SDSS, UKIDSS, and WISE. For each candidate, the templates are scaled to the observed photometry, and the best-fitting spectral type is assigned through $\chi^{2}$ minimization. To account for extragalactic contamination, quasar templates are also included, and a candidate is classified as a quasar if that fit yields a lower $\chi^{2}$ than an ultracool dwarf template. Applied to known ultracool dwarfs, the method achieves typical scatters of 1.5 subtypes for L dwarfs and 1.2 subtypes for T dwarfs. Its accuracy degrades toward fainter magnitudes ($J \sim 17.5$~mag), and unusually high $\chi^{2}$ values may indicate peculiar sources. This method, often combined with color cuts, has been applied or adapted to several surveys to generate large samples of ultracool dwarf candidates with photometric spectral types \citep[e.g.,][]{2016A&A...589A..49S, 2018ApJ...862..106T, 2019MNRAS.489.5301C, 2023MNRAS.522.1951D}. However, per-candidate template fitting is computationally intensive and not easily scalable to the data volumes of modern wide-field surveys. Furthermore, the use of a single averaged template for each subtype can bias the classifications of objects whose properties systematically deviate from typical ultracool dwarfs, such as blue, metal-poor subdwarfs or red, young L dwarfs.

In recent years, machine learning (ML) techniques have gained prominence in astronomy, valued for their scalability, flexibility, and ability to capture nonlinear patterns in complex datasets. Several studies have applied ML algorithms to ultracool dwarf searches. \cite{2019MNRAS.488.2263B} employed k-nearest neighbors \citep[KNN;][]{CoverHart1967} and simple feed-forward neural networks \citep{1986Natur.323..533R} to search for ultracool dwarfs, using 2MASS and AllWISE photometry across $\sim 40$~degree$^{2}$ sky regions. \cite{2022A&A...666A.147M} applied principal component analysis \citep[PCA;][]{Pearson1901, Jolliffe2002} and support vector machine \citep[SVM;][]{CortesVapnik1995} to photometry from the Javalambre Photometric Local Universe Survey \citep[J-PLUS;][]{2019A&A...622A.176C}. \cite{2022RNAAS...6...74G} and \cite{2022RNAAS...6...75G} applied random forests \citep[RF;][]{Breiman2001} to identify and classify ultracool dwarfs within $\sim 10$~degree$^{2}$ regions from SDSS+UKIDSS+WISE and Pan-STARRS1+2MASS+UKIDSS+AllWISE, respectively. \cite{2023PASP..135d4502S} tested KNN, RF, probabilistic random forest \citep[PRF;][]{2019AJ....157...16R}, and multilayer perceptron \citep[MLP;][]{Rosenblatt1958, Rumelhart1986, Ball2010} to estimate spectral types of M dwarfs using SDSS, 2MASS, and WISE photometry. More recently, \cite{2025MNRAS.541.1670B} applied KNN and artificial neural networks \citep[ANN;][]{1986Natur.323..533R} to identify late-T and Y dwarfs in a $\sim 1.5$~degree$^{2}$ region from the UKIDSS Ultra Deep Survey \citep{2007MNRAS.379.1599L}. 

While most of these studies trained ML models on spectroscopically confirmed M or ultracool dwarfs, \cite{2022RNAAS...6...74G} relied instead on photometric candidates from \cite{2016A&A...589A..49S}; also, \cite{2025MNRAS.541.1670B} used theoretical atmospheric models, which, as noted by the author, may systematically differ from bona fide ultracool dwarfs \citep[e.g.,][]{2021ApJ...916...53Z, 2021ApJ...921...95Z}. Additionally, most of the ML algorithms used in these works --- PCA, KNN, SVM, RF, MLP, and ANN --- do not natively handle missing values (i.e., non-detections in some bands). As a result, previous searches either (1) restricted their training samples and candidates to those detected in all required broadband filters \citep[][]{2019MNRAS.488.2263B, 2022A&A...666A.147M, 2022RNAAS...6...74G}, or (2) imputed missing values \citep[][]{2023PASP..135d4502S, 2025MNRAS.541.1670B}. The former approach excludes genuine ultracool dwarfs with incomplete photometry --- a common occurrence in wide-field searches --- while the latter risks introducing systematic biases.

Building on these developments, we present the Ultracool dwarf Science with MachIne LEarning (USMILE) program, which applies scalable and robust ML algorithms to (1) discover new ultracool dwarfs from wide-field surveys, such as LSST and Euclid, and (2) characterize these objects' physical properties. In this paper, we focus on the discovery of new ultracool dwarfs\footnote{Throughout this work, we define ultracool dwarfs as objects with spectral types of M6 or later.} and introduce the first version of the \texttt{USMILE} models, optimized for broadband photometry from LSST, VHS, and CatWISE, which we designate as \texttt{USMILE Avocado} for future reference. 

Our models are trained on a large, uncertainty-augmented labeled dataset (Section~\ref{sec:master_set}) that includes known ultracool dwarfs from the latest compilation, along with reddened early-type stars and quasars spanning a broad range of redshifts. Using gradient-boosted decision trees implemented with XGBoost \citep{2016arXiv160302754C}, we develop and validate two complementary models (Section~\ref{sec:method}): the \texttt{USMILE classifier}, which distinguishes ultracool dwarfs from contaminants, and the \texttt{USMILE regressor}, which predicts quantitative spectral types. These models are designed to efficiently scale to the massive data volumes of modern wide-field sky surveys. Another key advantage of XGBoost is its native ability to handle missing feature values without imputation --- a capability lacking in many alternative algorithms, such as PCA, SVM, KNN, RF, and ANN.

Applying these models, we carry out the first systematic ultracool dwarf search with LSST DP1, cross-matched against VHS Data Release 5 (DR5) and CatWISE2020 (Sections~\ref{sec:lsst_dp1}--\ref{sec:apply}). The overlap between LSST DP1 and Euclid within the EDF-S field provides a rare opportunity for large-scale external spectroscopic validation of our methodology (Section~\ref{sec:euclid}). This confirmed new ultracool dwarf discoveries, validated the performance of our \texttt{USMILE} models, and clarified regimes where \texttt{USMILE} predictions are most reliable. Guided by these results, we further identify additional high-quality photometric candidates of ultracool dwarfs that lack Euclid spectra and require future spectroscopic follow-up (Section~\ref{sec:final}). A summary of our findings is presented in Section~\ref{sec:summary}.

\section{Labeled Dataset}
\label{sec:master_set}
We first construct a comprehensive labeled dataset comprising ultracool dwarfs (Sections~\ref{subsec:ultracool} and \ref{subsec:latety}), reddened early-type stars, and quasars (Section~\ref{subsec:negative}). This dataset supports the training and testing of our ML models (Section~\ref{sec:method}). For each object, we compile or synthesize broadband photometry in the $\ilsst/\zlsst/\ylsst$ filters from LSST DP1 \citep{2025lsst.data....3N, 2025rubn.rept...31N},\footnote{In this work, we focus on the three longest-wavelength LSST filters --- $\ilsst$, $\zlsst$, and $\ylsst$ --- which are best suited to probing the red SEDs of ultracool dwarfs compared to $\ulsst$, $\glsst$, and $\rlsst$.} $\yvhs/\jvhs/\hvhs/\kvhs$ from VHS DR5 \citep{2013Msngr.154...35M}, and $\wa/\wb$ from CatWISE2020 \citep{2021ApJS..253....8M}. We then define eight color features using $\ylsst$ as the anchor: $\ilsst - \ylsst$, $\zlsst - \ylsst$, $\ylsst - \yvhs$, $\ylsst - \jvhs$, $\ylsst - \hvhs$, $\ylsst - \kvhs$, $\ylsst - \wa$, and $\ylsst - \wb$ (Section~\ref{subsec:augmentation}). Throughout this work, we adopt AB magnitudes for LSST and Vega magnitudes for VHS and CatWISE.

\subsection{Ultracool Dwarfs from UltracoolSheet}
\label{subsec:ultracool}

Our control sample of ultracool dwarfs is primarily drawn from UltracoolSheet Version 2.1.0 \citep[][]{ultracoolsheet}, released on July 3, 2025. This catalog compiles multi-wavelength photometry in the $\gps/\rps/\ips/\zps/\yps$ filters of Pan-STARRS1, the $\jmass/\hmass/\kmass$ filters of 2MASS, the $\ymko/\jmko/\hmko/\kmko$ filters based on the Mauna Kea Observatories (MKO) photometric system, and the $\wa/\wb$ filters from CatWISE.\footnote{When CatWISE photometry is unavailable, UltracoolSheet adopts $W1$ and $W2$ measurements from AllWISE. For consistency, we refer to all $W1$ and $W2$ photometry in UltracoolSheet as ``CatWISE'' throughout this work.} We exclude both unresolved and resolved binaries and require all objects to have (1) Pan-STARRS1 $y_{\rm P1}$-band photometry, and (2) spectral types of $\geqslant$M6, yielding a sample of 2,589 ultracool dwarfs. Although LSST photometry is not yet available and VHS data are largely missing for these selected objects, their available photometry from Pan-STARRS1, 2MASS, and MKO allows us to synthesize LSST (Section~\ref{subsubsec:ps2lsst}) and VHS (Section~\ref{subsubsec:vhs}) magnitudes. Such synthesis is feasible because the LSST $grizy$ filters closely match those of Pan-STARRS1, and the VHS $YJHK_{S}$ filters have similar wavelength coverage to those of 2MASS or MKO, ensuring that conversions can be derived with minimal extrapolation.

\begin{figure*}[t]
\begin{center}
\includegraphics[width=5.8in]{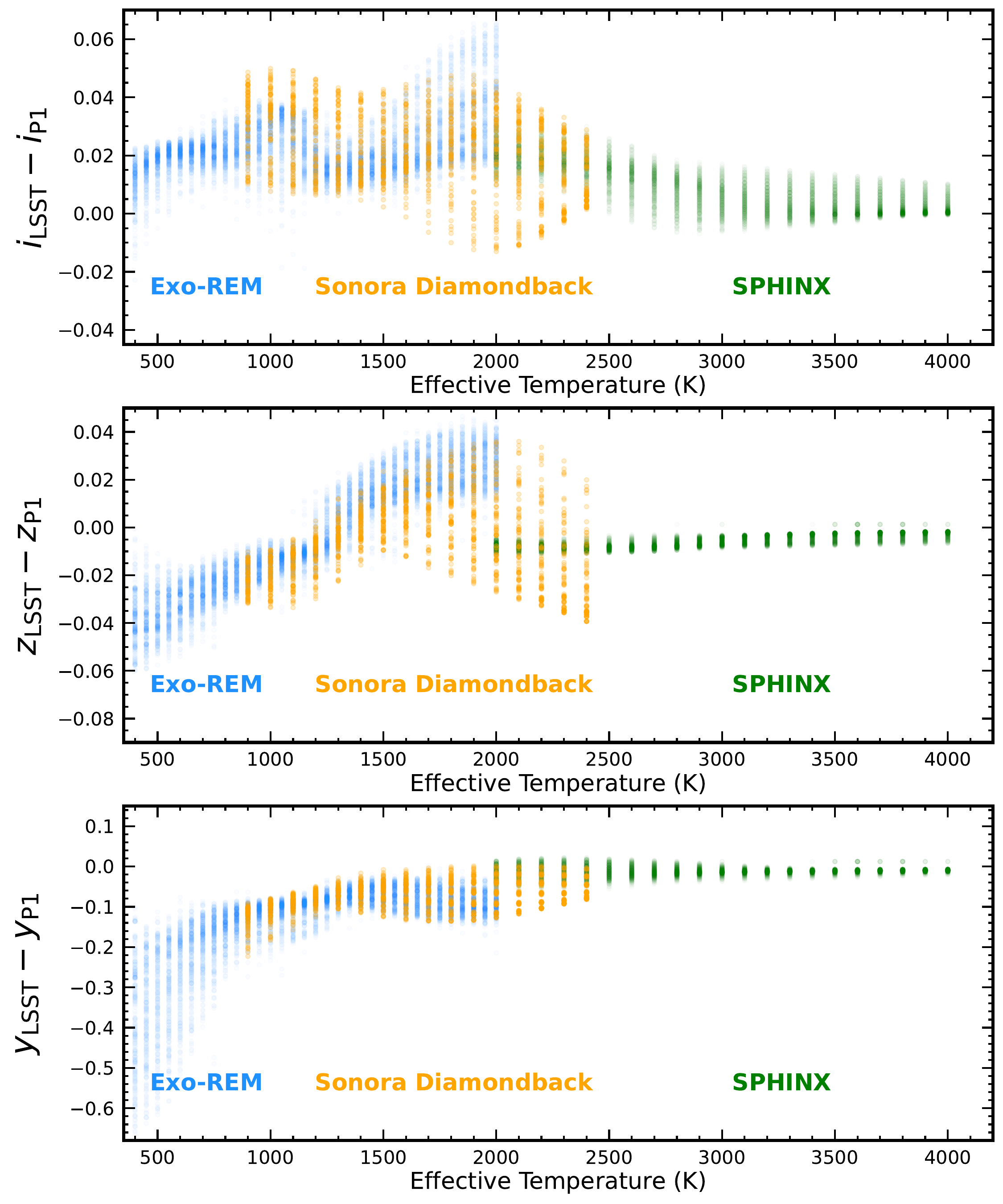}
\caption{Photometric differences between Pan-STARRS1 and LSST in the $i$ (top), $z$ (middle), and $y$ (bottom) bands, derived from synthetic model spectra of the \texttt{SPHINX} (green), \texttt{Sonora Diamondback} (orange), and \texttt{Exo-REM} (blue) grids. Each circle marks an individual synthetic spectrum from the respective model grid. }
\label{fig:ps2lsst}
\end{center}
\end{figure*}

\subsubsection{Synthesizing LSST Photometry from Pan-STARRS1}
\label{subsubsec:ps2lsst}

Photometric conversions between filter systems are usually derived from direct comparisons of observed magnitudes across surveys. Such empirical conversions are not yet feasible for LSST, given its early phase of sky coverage. Spectroscopy of ultracool dwarfs can, in principle, be used to compute synthetic magnitudes, but most available near-infrared spectra extend only to $\sim 0.8$~$\mu$m at the shortest-wavelength end and suffer from growing flux uncertainties toward the detector edge \citep[e.g.,][]{2003ApJ...596..561M, 2007ApJ...658.1217M, 2014ASInC..11....7B, 2015ApJ...814..118B, 2017ApJ...837...95B, 2015ApJ...810..158F, 2015ApJS..219...33G, 2018AJ....155...34C, 2018ApJ...858...41Z, 2021ApJ...921...95Z, 2022ApJ...935...15Z, 2021ApJS..257...45H, 2024ApJS..274...40H, 2023ApJ...959...63S, 2024ApJ...961..121H}. Optical spectral libraries for ultracool dwarfs remain sparse compared to near-infrared ones for these intrinsically red ultracool dwarfs \citep[though see, e.g.,][]{1999ApJ...519..802K, 2000AJ....120..447K, 2010AJ....139.1808S, 2015AJ....149..158S, 2011AJ....141...97W}. As a consequence, direct spectral synthesis of LSST magnitudes is generally inaccessible. In this context, theoretical models of ultracool dwarf atmospheres offer a practical alternative for deriving photometric conversions between Pan-STARRS1 and LSST. 

To establish these conversions, we employ three grids of atmospheric models, each representing different assumptions about ultracool dwarf atmospheres:
\begin{enumerate}
\item[$\bullet$] \texttt{SPHINX} \citep{2023ApJ...944...41I}, which assumes cloud-free atmospheres in chemical equilibrium. They span effective temperatures ($T_{\rm eff}$) from 2000~K to 4000~K, logarithmic surface gravities ($\log{(g)}$) from 4~dex to 5.5~dex, metallicities ([M/H]) from $-1$~dex to $+1$~dex, and carbon-to-oxygen ratios (C/O) from 0.3 to 0.9.
\item[$\bullet$] \texttt{Sonora Diamondback} \citep{2024ApJ...975...59M}. These models assume cloudy atmospheres in chemical equilibrium, with $T_{\rm eff} \in$[900~K, 2400~K], $\log{(g)} \in$[3.5~dex, 5.5~dex], and [M/H]$\in$[$-0.5$~dex, $+0.5$~dex]. They also include a parameter, $f_{\rm sed}$, which characterizes the cloud sedimentation efficiency. These models are available at six $f_{\rm sed}$ values: 1, 2, 3, 4, 8, ``nc'', where ``nc'' denotes a cloud-free atmosphere.
\item[$\bullet$] \texttt{Exo-REM} \citep{2015A&A...582A..83B, 2018ApJ...854..172C, 2021A&A...646A..15B}. These models assume cloudy atmospheres in chemical disequilibrium, with $T_{\rm eff} \in$[400~K, 2000~K], $\log{(g)} \in$[3~dex, 5~dex], [M/H]$\in$[$-0.5$~dex, $+1$~dex], and C/O$\in$[0.1, 0.8].
\end{enumerate}
For each model spectrum, we synthesize Pan-STARRS1 and LSST magnitudes,\footnote{Filter transmission curves and zero-point fluxes are obtained from the SVO Filter Profile Service: \url{http://svo2.cab.inta-csic.es/svo/theory/fps/index.php}. The curves account for the filter, instrument, and Earth's atmospheric transmission. MKO filter profiles are from \cite{2006MNRAS.367..454H}; others are taken from the corresponding instrument webpages. } and then compute magnitude differences: $\Delta{i} = \ilsst - \ips$, $\Delta{z} = \zlsst - \zps$, and $\Delta{y} =  \ylsst - \yps$. 

As shown in Figure~\ref{fig:ps2lsst}, the model-predicted offsets between Pan-STARRS1 and LSST are generally $<0.1$~mag across a wide range of $T_{\rm eff}$, except for $\Delta{y}$ at $T_{\rm eff} < 1200$~K. Although different models predict slightly different median values and scatter, the results are broadly consistent. To reduce model dependence, we adopt the mean magnitude differences of all available model predictions as the final conversion factor, with details provided below.

For objects with $T_{\rm eff}$ and $\log{g}$ estimates from \citet{2023ApJ...959...63S}, we compile $\Delta{i}$, $\Delta{z}$, and $\Delta{y}$ values from all models with $T_{\rm eff}$ within $\pm 150$~K and $\log{g}$ within $\pm 0.5$~dex of the reported parameters.\footnote{The average uncertainties in $T_{\rm eff}$ and $\log{(g)}$ reported by \cite{2023ApJ...959...63S} are 117~K and 0.34~dex, respectively. Our adopted ranges of $\pm 150$~K and $\pm 0.5$~dex are chosen to encompass these typical errors.} For each model grid, we compute the mean and standard deviation of the $\Delta$ magnitudes, average the means across all available grids, propagate uncertainties, and adopt the result as the final conversion factor between Pan-STARRS1 and LSST.

The remaining ultracool dwarfs not analyzed by \cite{2023ApJ...959...63S} consist of a mixture of subdwarfs and normal (young or field-age) dwarfs. For subdwarfs with M6--L7 spectral types, we estimate $T_{\rm eff}$ using the empirical spectral type--$T_{\rm eff}$ relation of \cite{2017MNRAS.464.3040Z}. We then compute conversion factors by collecting $\Delta{i}$, $\Delta{z}$, and $\Delta{y}$ from all available model grids within $\pm 150$~K of the estimated $T_{\rm eff}$ and [M/H]$\leqslant 0$~dex. For subdwarfs later than L7, we adopt the field-age spectral type--$T_{\rm eff}$ relation from \cite{2023ApJ...959...63S} and allow a broader $T_{\rm eff}$ window of $\pm 300$~K, again using only model predictions with [M/H]$\leqslant 0$~dex. Given that the empirical relation of \cite{2023ApJ...959...63S} was derived with subdwarfs excluded, the broader $T_{\rm eff}$ range of $\pm 300$~K in our analysis accounts for the potential offsets in $T_{\rm eff}$ between subdwarfs and normal dwarfs at late-L and T types, as observed for late-M and L subdwarfs \citep[e.g.,][]{2006ApJ...645.1485B, 2017MNRAS.464.3040Z}.

For the remaining normal ultracool dwarfs, we estimate $T_{\rm eff}$ using the spectral type--$T_{\rm eff}$ relations from \cite{2023ApJ...959...63S}, applying the young and field-age relations for objects younger or older than 300~Myr, respectively. We then derive photometric conversion factors from all model predictions with $T_{\rm eff}$ within $\pm150$~K of the estimated value. 

Finally, we compute $\ilsst$, $\zlsst$, and $\ylsst$ magnitudes for all ultracool dwarfs by adding the corresponding $\Delta$ magnitudes to their Pan-STARRS1 photometry, with uncertainties propagated. To minimize model dependence, we perform the conversions strictly within the same band rather than across different bands. Consequently, the completeness (or, in other words, fraction of missing values) of synthesized LSST magnitudes in each band directly inherits that of the original Pan-STARRS1 data.

\subsubsection{Synthesizing VHS Photometry from 2MASS and MKO}
\label{subsubsec:vhs}

The VHS $\yvhs$, $\jvhs$, and $\hvhs$ filters have response curves similar to those of the MKO system, while the VHS $\kvhs$ filter more closely resembles 2MASS. Accordingly, UltracoolSheet incorporates VHS photometry for a small subset of objects, when available, within its 2MASS and MKO photometry columns. For these objects, we directly adopt the VHS photometry provided by UltracoolSheet. 

For the remaining ultracool dwarfs, we focus on objects with published IRTF/SpeX near-infrared spectra \citep{2004ApJ...614L..73B, 2007ApJ...659..655B, 2004AJ....127.2856B, 2006ApJ...639.1095B, 2006ApJ...637.1067B, 2007ApJ...658..557B, 2008ApJ...681..579B, 2009ApJ...697..148B, 2010ApJ...710.1142B, 
2006AJ....131.1007B, 2006ApJ...645.1485B, 2017ASInC..14....7B, 
2004ApJ...604L..61C, 
2006AJ....131.2722C, 
2006AJ....132.2074M, 
2006ApJ...639.1114R, 2006AJ....132..891R, 
2007ApJ...655..522L, 
2007AJ....134.1162L, 2008ApJ...686..528L, 
2007ApJ...654..570L, 2012ApJ...760..152L, 2014ApJ...787..126L, 
2007AJ....133.2320S, 
2008ApJ...676.1281M, 
2009AJ....137..304S, 
2010ApJ...715..561A, 
2010AJ....139..176F, 2016ApJS..225...10F, 
2011AJ....142...77D, 2012ApJ...755...94D, 2012ApJ...757..100D, 2014ApJ...792..119D, 2017MNRAS.467.1126D, 
2011ApJ...732...56G, 
2011ApJ...736L..34G, 2012AJ....144...94G, 2011AJ....142..171G, 
2010ApJS..190..100K, 2011ApJS..197...19K, 2014ApJ...783..122K, 
2012ApJS..201...19D, 
2013ApJ...772...79A, 
2013ApJ...776..126C, 
2013ApJ...777L..20L, 2016ApJ...833...96L, 
2013ApJS..205....6M, 
2013PASP..125..809T, 
2014ApJ...794..143B,
2013ApJ...777...84B, 2015ApJ...814..118B, 2017ApJ...837...95B,
2014ApJ...784...65B, 
2015ApJS..219...33G, 
2016AJ....151...46A,
2016ApJ...821..120A,
2016PhDT.......189A,
2016Natur.533..221G, 
2017AJ....154..112K, 
2021ApJ...911....7Z, 2021ApJ...921...95Z,
2023ApJ...959...63S, 
2024ApJ...961..121H}. Then we compute transformations between 2MASS, MKO, and VHS magnitudes through synthetic photometry. For the $Y$, $J$, and $H$ bands, we prioritize conversions based on MKO photometry, using 2MASS only when MKO magnitudes are unavailable. For the $K_{S}$ band, we reverse the order, prioritizing 2MASS. As with the  LSST photometry, we perform conversions only within the same band to preserve the completeness of the original 2MASS and MKO photometry.

\subsection{Additional Ultracool Dwarfs Near the T/Y Transition}
\label{subsec:latety}

We also include 19 of the 20 late-T and Y dwarfs from \citet{2024ApJ...976...82T}, all observed with JWST NIRSpec and MIRI spanning 0.95--12~$\mu$m in wavelength. The excluded object, WISE~J220905.73+271143.9 (Y0), has a low-quality NIRSpec spectrum with signal-to-noise ratios (S/Ns) $<2$ at wavelengths shorter than 2.5~$\mu$m. As noted by \citet{2024ApJ...976...82T}, these spectra of 20 objects were originally obtained through the JWST program GO 2302 (PI: M. C. Cushing) for 19 targets and the program GTO 1189 (PI: T. L. Roellig) for the remaining one target, 2MASS~J03480772-6022270. We note that the GO 2302 spectra were first published and analyzed by \cite{2024ApJ...973..107B}.

From these JWST spectra, we synthesize $\jvhs$, $\hvhs$, $\kvhs$, $\wa$, and $\wb$ magnitudes. We also compute $\ylsst$ magnitudes, as this band serves as the anchor for features in the labeled dataset (Section~\ref{subsec:augmentation}). Because the NIRSpec spectra only partially cover the $\ylsst$ bandpass, we linearly extrapolate the flux shortward of the spectrum to zero at zero wavelength to complete the coverage. Given the similar wavelength coverage of the $\yvhs$ and $\ylsst$ filters, we additionally synthesize $\yvhs$ magnitudes as well. The $\ilsst$ and $\zlsst$ magnitudes are left as missing values.

Additionally, we note that the synthesized photometry from the spectra of these late-T and Y dwarfs, as well as the negative samples described in Section~\ref{subsec:negative}, can vary in absolute scale depending on whether the spectra were normalized and on the objects’ distances. These variations in the absolute values of synthesized magnitudes do not affect our analysis, because we use only colors --- which are distance-independent --- and our machine-learning models are trained and tested exclusively on these color features (Section~\ref{subsec:augmentation}).

\subsection{Negative Samples}
\label{subsec:negative}

Our negative samples include reddened early-type stars and quasars. To assemble the stellar contaminants, we compile optical and near-infrared spectra for 689 O5--M5 giant and dwarf stars from the Pickles Spectral Library \citep[PSL;][]{1998PASP..110..863P} and the X-shooter Spectral Library \citep[XSL;][]{2022A&A...660A..34V}. For XSL, we retain only objects with labeled spectral types and spectra that have been corrected for both Galactic extinction and slit flux losses. The PSL and XSL spectra extend from 0.1~$\mu$m (PSL) or 0.35~$\mu$m (XSL) to 2.5~$\mu$m in wavelength, providing sufficient wavelength coverage to synthesize LSST and VHS photometry. We extend each stellar spectrum with a Rayleigh-Jeans tail to compute $\wa$ and $\wb$ photometry.

For extragalactic contaminants, we use the SWIRE Template Library \citep{2007ApJ...663...81P}, focusing on type-1 and type-2 quasar templates. Other galaxy types are excluded since those extended sources may be filtered out with point-source flags from LSST (e.g., \texttt{y\_extendedness}) and other catalogs. Quasar templates are redshifted over a grid from $z = 0$ to 6 in steps of 0.1. At $z=0$, these spectra span 0.1--6000~$\mu$m, enabling synthesis of all LSST, VHS, and CatWISE photometry of interest. At higher redshifts, $\ilsst$ and $\zlsst$ fall out of coverage around $z \approx 4$ and $z \approx 5$, respectively, so these bands are flagged as missing for high-$z$ quasars. The maximum redshift of $z=6$ ensures that $\ylsst$ photometry --- used as the anchor to define all color features for the labeled dataset --- remains available.

We further apply interstellar reddening to all stellar and quasar spectra using a grid of $V$-band extinctions from $A_{V} = 0$~mag to 10~mag in steps of 1~mag. We adopt the \cite{2011ApJ...737..103S} extinction law with $R_{V}$ fixed at 3.1. These procedures yield a total of 12,595 stellar and extragalactic contaminants.

\begin{figure}[t]
\begin{center}
\includegraphics[width=3.3in]{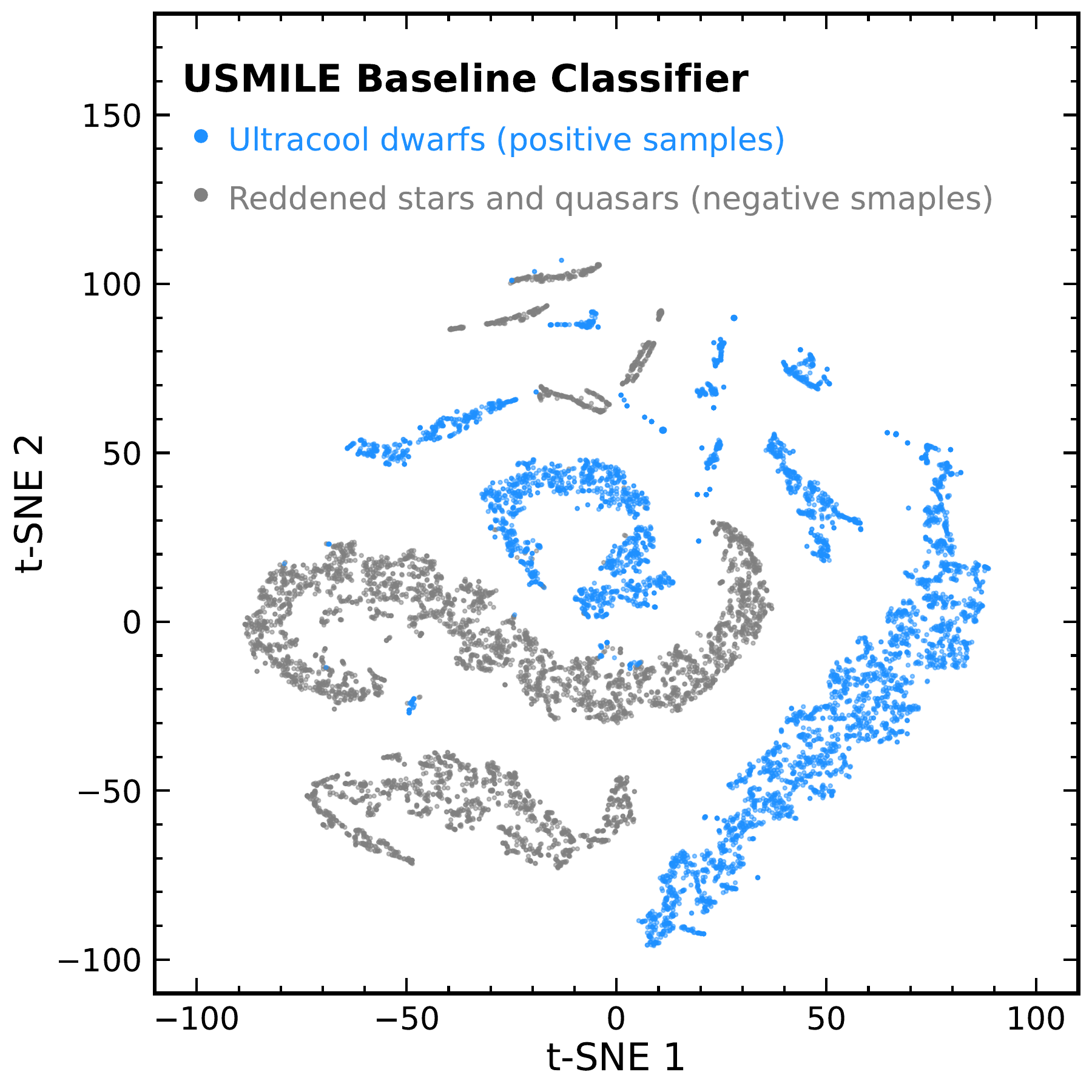}
\caption{The t-SNE projection of the complete labeled dataset (Section~\ref{subsec:classifier}). Positive (ultracool dwarfs) and negative (reddened early-type stars and quasars) samples are shown in blue and grey, respectively. }
\label{fig:tsne}
\end{center}
\end{figure}

\begin{figure*}[t]
\begin{center}
\includegraphics[width=6.5in]{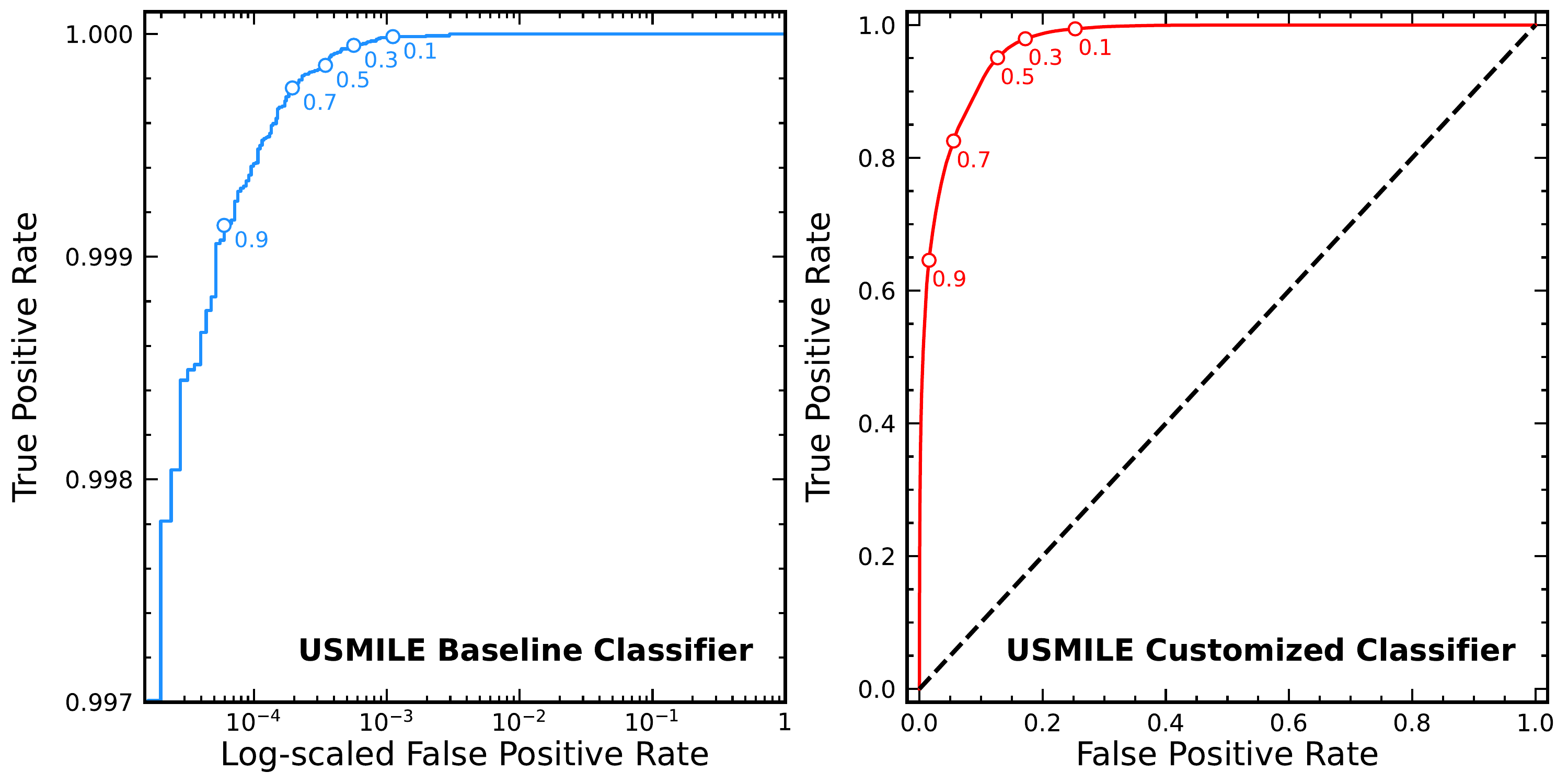}
\caption{ROC curves for the \texttt{USMILE baseline classifier} (left) and \texttt{customized classifier} (right). The corresponding area-under-the-curve (AUC) values are summarized in Table~\ref{tab:performance}. Open circles mark the points corresponding to the selected probability threshold for binary classification. In the left panel, the false-positive rate on the x-axis is shown on a logarithmic scale. }
\label{fig:roc}
\end{center}
\end{figure*}

\subsection{Augmentation and Assembly}
\label{subsec:augmentation}

To propagate photometric uncertainties, we augment the labeled dataset by generating 500 realizations for each ultracool dwarf and 100 realizations for each negative sample. For each realization, magnitudes are drawn from Gaussian distributions centered on the measured or synthesized values in a given band, with standard deviations equal to the associated photometric uncertainties. For late-T and Y dwarfs from \cite{2024ApJ...976...82T} and for negative samples, we adopt mean photometric uncertainties of ultracool dwarfs drawn from UltracoolSheet (Section~\ref{subsec:ultracool}): $\sigma =$0.022, 0.021, 0.037, 0.053, 0.019, 0.033, 0.054, 0.015, 0.015~mag in the $\ilsst$, $\zlsst$, $\ylsst$, $\yvhs$, $\jvhs$, $\hvhs$, $\kvhs$, $\wa$, and $\wb$ bands, respectively. The unequal numbers of realizations for positive (ultracool dwarfs) and negative samples are chosen to maintain class balance. 

Finally, we define eight color features anchored on $\ylsst$: $\ilsst - \ylsst$, $\zlsst - \ylsst$, $\ylsst - \yvhs$, $\ylsst - \jvhs$, $\ylsst - \hvhs$, $\ylsst - \kvhs$, $\ylsst - \wa$, and $\ylsst - \wb$. All these features are independent of the objects' distances from Earth. We assign ``Y'' labels to genuine ultracool dwarf and ``N'' labels to negative samples. 

The resulting labeled dataset contains 2,541,500 objects, evenly split between ultracool dwarfs and contaminants. Table~\ref{tab:missing} summarizes the fraction of missing values in each feature. This dataset provides the basis for training and testing the \texttt{USMILE} tree-based models, described in the following section.

\renewcommand{\arraystretch}{1.25} 
{ 
\begin{deluxetable*}{lccccc}
\setlength{\tabcolsep}{10pt} 
\tablecaption{Fractions of Missing Feature Values in the Labeled Dataset and Initial LSST+VHS+CatWISE Candidates} \label{tab:missing} 
\tablehead{ \multicolumn{1}{l}{}   &   \multicolumn{2}{c}{Labeled Dataset (2,541,500)}   &   \multicolumn{1}{c}{}   &   \multicolumn{2}{c}{Initial Candidates (4,053)}    \\ 
\cline{2-3} \cline{5-6} 
\multicolumn{1}{l}{Property}   &   \multicolumn{1}{c}{Number of missing values}   &   \multicolumn{1}{c}{Fraction}   &   \multicolumn{1}{c}{}   &   \multicolumn{1}{c}{Number of missing values}   &   \multicolumn{1}{c}{Fraction}    } 
\ 
\startdata 
$i_{\rm LSST} - y_{\rm LSST}$ & 445,800 & 17.5\% &   & 1,732 & 42.7\% \\ 
$z_{\rm LSST} - y_{\rm LSST}$ & 141,200 & 5.6\% &   & 1,897 & 46.8\% \\ 
$y_{\rm LSST} - Y_{\rm MKO}$ & 787,500 & 31.0\% &   & 3,980 & 98.2\% \\ 
$y_{\rm LSST} - J_{\rm VHS}$ & 779,500 & 30.7\% &   & 3,277 & 80.9\% \\ 
$y_{\rm LSST} - H_{\rm VHS}$ & 780,000 & 30.7\% &   & 3,730 & 92.0\% \\ 
$y_{\rm LSST} - K_{S,\rm VHS}$ & 833,000 & 32.8\% &   & 3,638 & 89.8\% \\ 
$y_{\rm LSST} - W1_{\rm CatWISE}$ & 16,000 & 0.6\% &   & 2,233 & 55.1\% \\ 
$y_{\rm LSST} - W2_{\rm CatWISE}$ & 18,000 & 0.7\% &   & 2,599 & 64.1\% \\ 
Spectral Type & 700,700 & 27.6\% &   & -- & -- \\ 
\enddata 
\end{deluxetable*} 
}

\begin{figure*}[t]
\begin{center}
\includegraphics[width=7in]{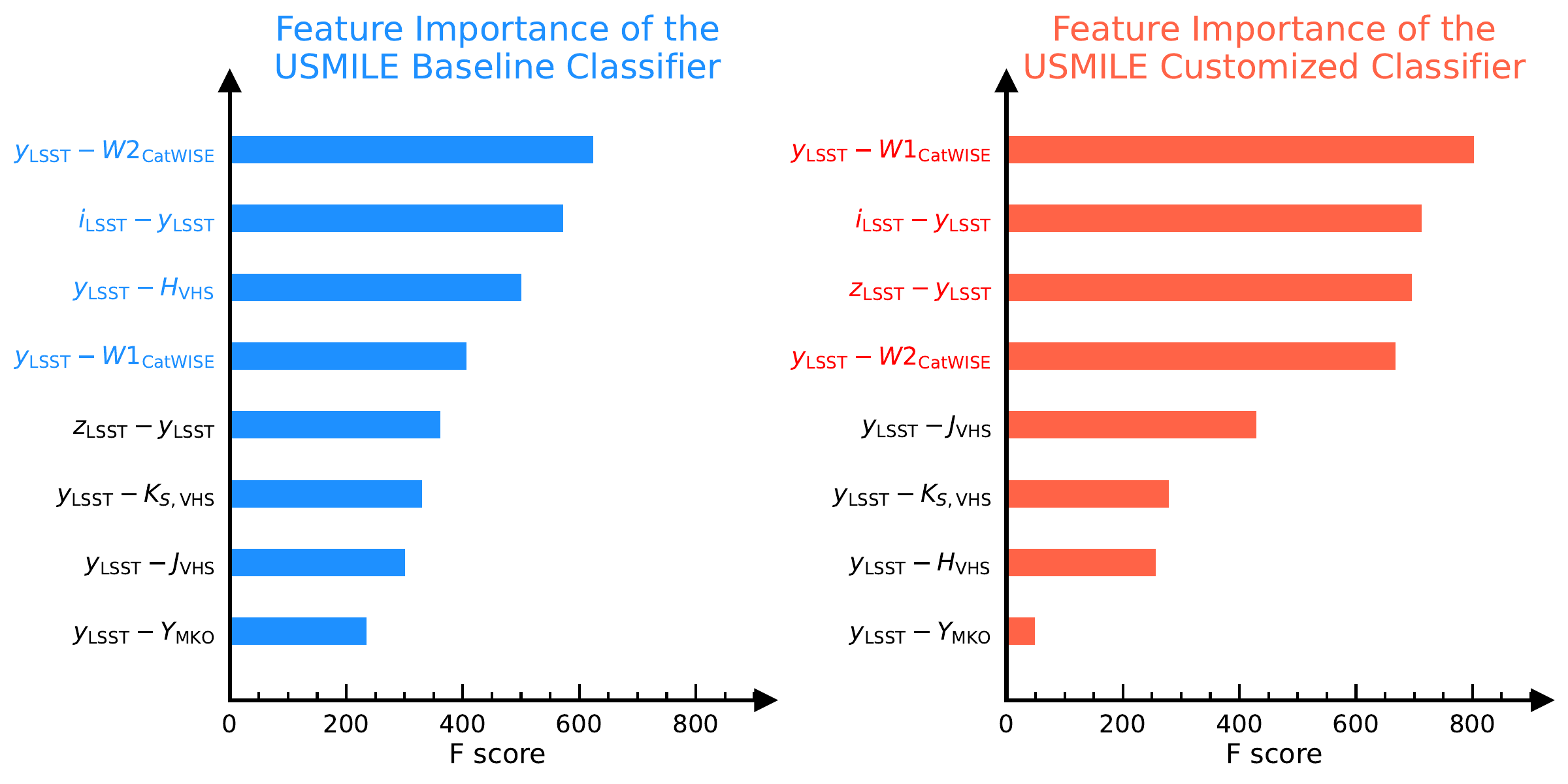}
\caption{Feature importance for the \texttt{USMILE baseline classifier} (left) and \texttt{customized classifier} (right), ranked by XGBoost F scores (i.e., the frequency with which each feature is used to split the data across all trees). The four most influential features for each classifier are highlighted in color.}
\label{fig:feature_important_classifier}
\end{center}
\end{figure*}

\begin{figure*}[t]
\begin{center}
\includegraphics[width=6.5in]{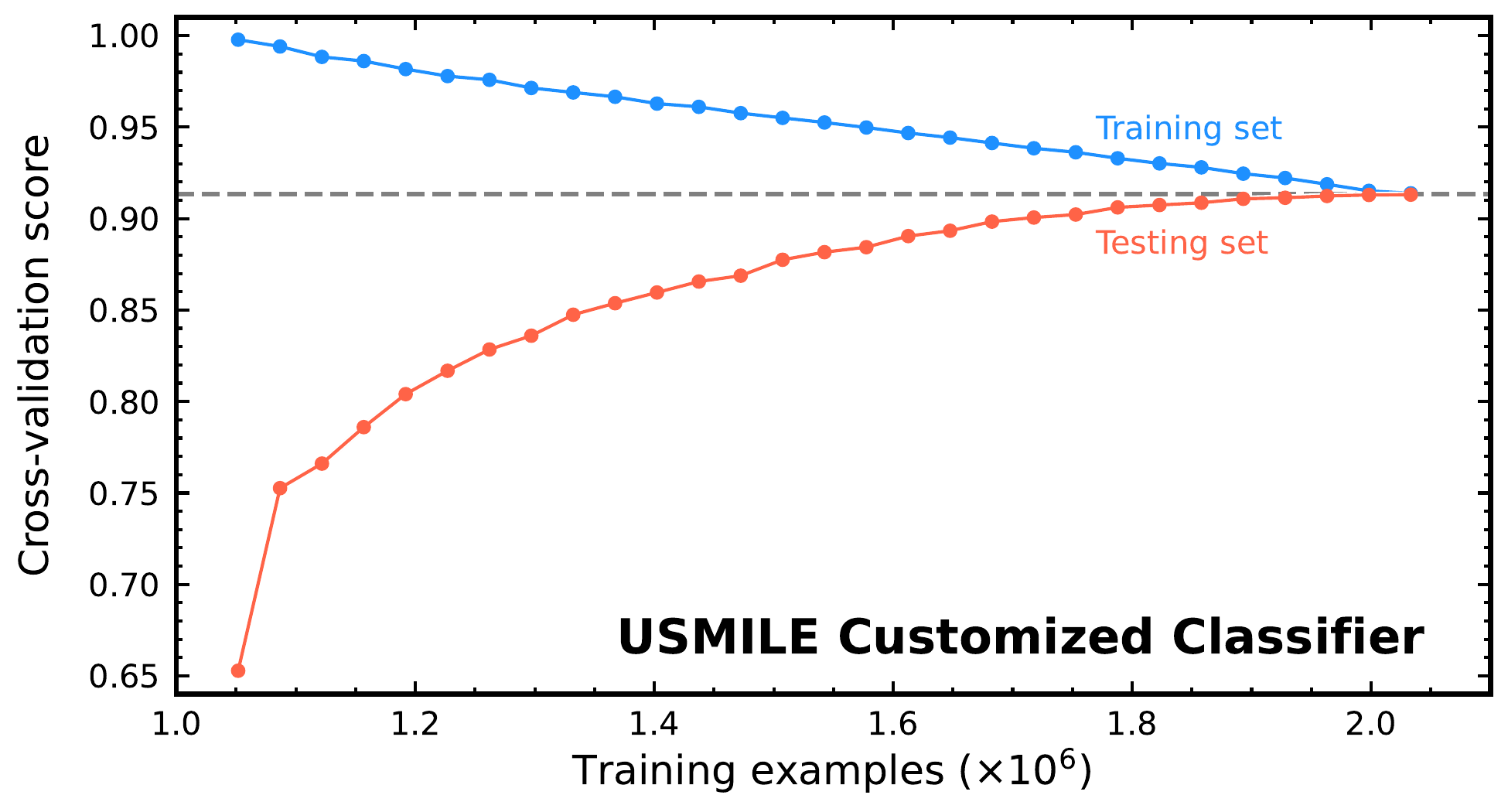}
\caption{Learning curves of a representative \texttt{USMILE customized classifier}, showing cross-validation scores for the training set (blue) and testing set (red). These scores are computed using 5-fold stratified cross-validation with shuffling to preserve class balance across folds. Both curves converge at large sample sizes near 0.91, demonstrating the model's strong generalization performance with a well-balanced bias--variance trade-off. }
\label{fig:learning_curve_classifier}
\end{center}
\end{figure*}

\begin{figure*}[t]
\begin{center}
\includegraphics[width=7.in]{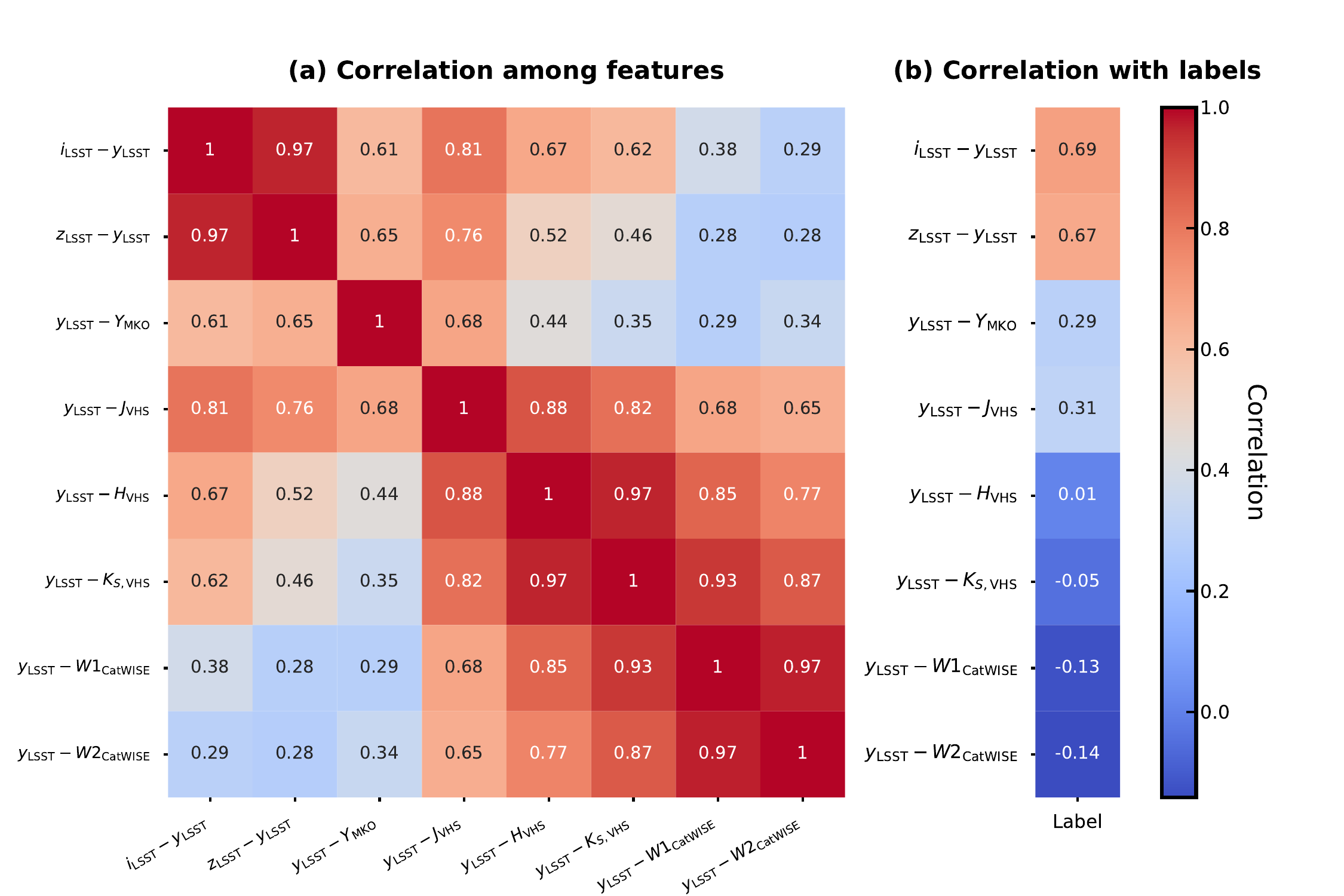}
\caption{Linear correlations among the eight color features (left) and between each feature and the labels (right), where labels indicate whether a source is a genuine ultracool dwarf or a stellar/extragalactic contaminant. Pearson correlation coefficients are labeled. }
\label{fig:classifier_corr}
\end{center}
\end{figure*}

\section{The \texttt{USMILE} Tree-based Models}
\label{sec:method}

Tree-based models are non-parametric supervised learning algorithms that recursively partition the feature space into subregions with relatively homogeneous target values, constructing decision trees to model non-linear relationships and account for heterogeneous feature types. While individual decision trees are straightforward to interpret, they often suffer from high variance and limited predictive stability. Ensemble approaches such as gradient-boosted decision trees \citep[GBDT;][]{Friedman2001} overcome these limitations by combining multiple weak learners into a strong predictor. In GBDT, trees are built sequentially, with each new tree modeling the residuals of the previous ensemble to minimize a differentiable loss function.

For this work, we employ XGBoost \citep{2016arXiv160302754C}, an industry-standard implementation of GBDT that combines algorithmic improvements with system-level optimizations. Notably, its sparsity-aware split-finding algorithm assigns default directions at tree nodes and thereby natively routes missing feature values without the need for imputation. This capability --- absent in many alternative algorithms (Section~\ref{sec:introduction}) --- is critical for ultracool dwarf surveys, where non-detections are common and could otherwise lead to data loss or systematic bias. Furthermore, XGBoost incorporates cache-aware access patterns and parallel learning, making it highly scalable to the massive datasets in modern wide-field sky surveys.

We develop two complementary models with this framework, trained on the labeled dataset described in Section~\ref{sec:master_set}: the \texttt{USMILE classifier} (Section~\ref{subsec:classifier}) and \texttt{regressor} (Section~\ref{subsec:regressor}). The \texttt{classifier} performs a binary classification to determine whether a candidate is an ultracool dwarf, while the \texttt{regressor} provides a quantitative estimate of spectral types for classified ultracool dwarfs. Together, these models form an end-to-end pipeline tailored to the discovery and characterization of ultracool dwarfs in large-scale sky surveys.

\subsection{USMILE Classifier}
\label{subsec:classifier}

We begin by assessing the separability of the labeled dataset using t-distributed Stochastic Neighbor Embedding \citep[t-SNE;][]{vanDerMaaten2008}. As shown in Figure~\ref{fig:tsne}, projecting all eight color features into two t-SNE dimensions reveals that $\geqslant$M6 ultracool dwarfs (positive samples) form distinct clusters that are well-separated from reddened early-type stars and quasars (negative samples). This pronounced separation demonstrates that the eight color features adopted in this work contain strong discriminative power for training high-accuracy classifiers.

To leverage the full predictive potential of the labeled dataset, we first train a \texttt{baseline classifier} using XGBoost's implementation of a gradient-boosted decision tree classifier. The labeled dataset is randomly split into training and testing sets in an 80:20 ratio. The ensemble consists of 100 decision trees with a maximum depth of 6 levels and a learning rate of 0.3. A ridge regularization term is applied to leaf weights to mitigate overfitting, and all available features are considered at each split. The model is optimized for binary classification with a logarithmic loss function. The predicted probabilities by the \texttt{classifier} are converted to binary labels (i.e., ultracool dwarf vs. contaminant) using a threshold of 0.5, chosen based on Youden's index on the Receiver Operating Characteristic (ROC) curve.

The \texttt{baseline classifier} achieves strong predictive performance. Its ROC curve (Figure~\ref{fig:roc}) yields an area under the curve (AUC) close to 1.0, consistent with the clear separation seen in the t-SNE projection (Figure~\ref{fig:tsne}). Table~\ref{tab:performance} further summarizes multiple evaluation metrics on both training and testing sets, showing that the model generalizes effectively with no evidence of overfitting or underfitting. According to Figure~\ref{fig:feature_important_classifier}, the most important features for the \texttt{baseline classifier} are $\ylsst-\wb$, $\ilsst-\ylsst$, $\ylsst-\hvhs$, and $\ylsst-\wa$, underscoring the value of combining LSST, VHS, and CatWISE data in ultracool dwarf searches.

When applying the classifier to candidates identified in our initial search from LSST DP1, VHS, and CatWISE (see Section~\ref{sec:lsst_dp1}), we find substantial differences in the fraction of missing feature values between the labeled dataset and the initial candidate list. As shown in Table~\ref{tab:missing}, the color features of candidates are overall much sparser. To reduce potential biases and maintain robust model performance, we train \texttt{customized classifiers} tailored to these missing-data patterns. For each feature, we randomly mask values in the labeled dataset until the overall missing-value fraction matches that of the candidates, and then retrain the classifier using the same procedure as the \texttt{baseline classifier}. This process is repeated 400 times, producing 400 \texttt{customized classifiers}.

The performance of these \texttt{customized classifiers} is only slightly below that of the \texttt{baseline classifier}, remaining consistently high. A typical \texttt{customized classifier} achieves a ROC AUC of 0.97 (Figure~\ref{fig:roc}), with accuracy of 0.91, precision of 0.88, recall of 0.95, and F1 score of 0.92 (Table~\ref{tab:performance}). Performance variation across all 400 classifiers is $<0.01$ for each metric. As shown in Figure~\ref{fig:learning_curve_classifier}, the learning curves of both training and testing sets converge, with cross-validation scores stabilizing around 0.91. These results confirm that the \texttt{classifier} generalizes well with a balanced bias--variance trade-off.

As shown in Figure~\ref{fig:feature_important_classifier}, the most influential features for the \texttt{customized classifiers} are $\ylsst-\wa$, $\ilsst-\ylsst$, $\zlsst-\ylsst$, and $\ylsst-\wb$. The reduced importance of VHS-based colors relative to the \texttt{baseline classifiers} likely reflects the high fraction of missing VHS data among the initial candidates (Table~\ref{tab:missing}). 

We also examine linear correlations among features and between features and labels for the \texttt{customized classifiers} (Figure~\ref{fig:classifier_corr}). Several color pairs exhibit strong mutual correlations, including (1) $\ilsst-\ylsst$ with $\zlsst-\ylsst$, (2) $\ylsst-\hvhs$ with $\ylsst-\kvhs$, and (3) $\ylsst-\wa$ with $\ylsst-\wb$, each with a linear correlation coefficient of 0.97. These high inter-feature correlations likely arise from the adjacent wavelength coverage of the corresponding filters (e.g., $\ilsst$ and $\zlsst$, $\hvhs$ and $\kvhs$, $\wa$ and $\wb$), which capture similar portions of an object's SED. In addition, $\ilsst-\ylsst$ and $\zlsst-\ylsst$ show strong positive correlations with the labels (0.69 and 0.67, respectively), consistent with expectations that redder optical colors trace cooler effective temperatures and thus identify ultracool dwarfs. 

\begin{figure*}[t]
\begin{center}
\includegraphics[width=7.in]{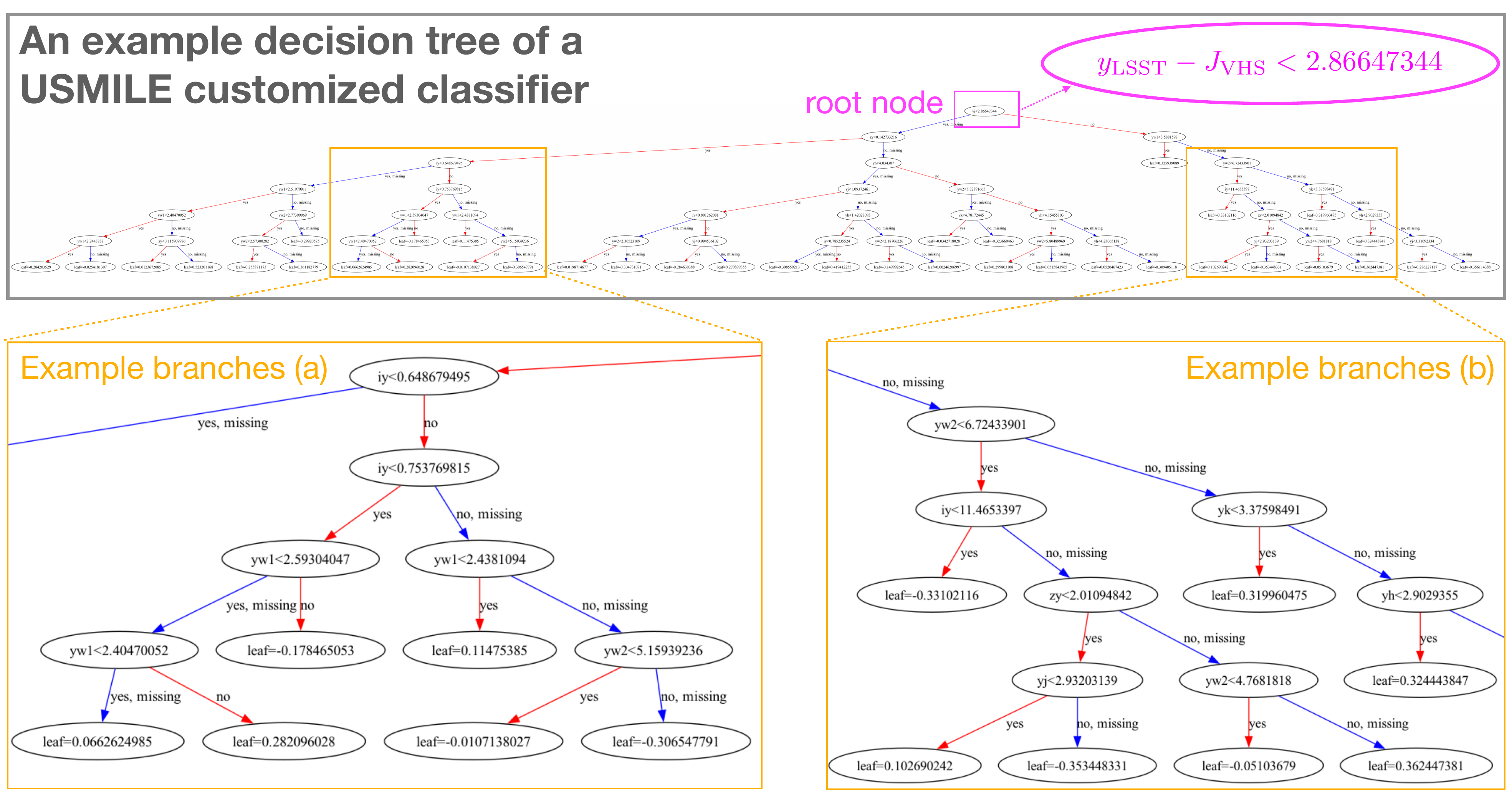}
\caption{Visualization of the 30th decision tree from a \texttt{USMILE customized classifier}; each classifier contains 100 decision trees, with 400 \texttt{classifiers} trained in total. Each split evaluates a single color feature against a threshold, branching into two possible paths: a default direction (blue) and an alternate direction (red). Two example regions of the tree are shown, where ``iy'', ``zy'', ``yj'', ``yh'', ``yk'', ``yw1'', and ``yw2'' correspond to the $\ilsst-\ylsst$, $\zlsst-\ylsst$, $\ylsst-\jvhs$, $\ylsst-\hvhs$, $\ylsst-\kvhs$, $\ylsst-\wa$, and $\ylsst-\wb$ colors, respectively. The resulting log-odds scores of the leaf nodes are then combined to yield the final classification probability.  }
\label{fig:tree}
\end{center}
\end{figure*}

Finally, we illustrate the decision-making process by visualizing one of 100 decision trees from a \texttt{customized classifier} in Figure~\ref{fig:tree}. In this example, the root node evaluates $\ylsst-\jvhs$, followed by hierarchical splits on the same or other color features. The same feature may appear multiple times with different thresholds, and root nodes vary across trees. Each split corresponds to a simple color cut, a method long used in ultracool dwarf searches (Section~\ref{sec:introduction}). By embedding these cuts in a recursive, data-driven, and scalable framework, machine learning generalizes traditional techniques into a powerful classifier with high predictive accuracy.

\begin{figure*}[t]
\begin{center}
\includegraphics[width=7.in]{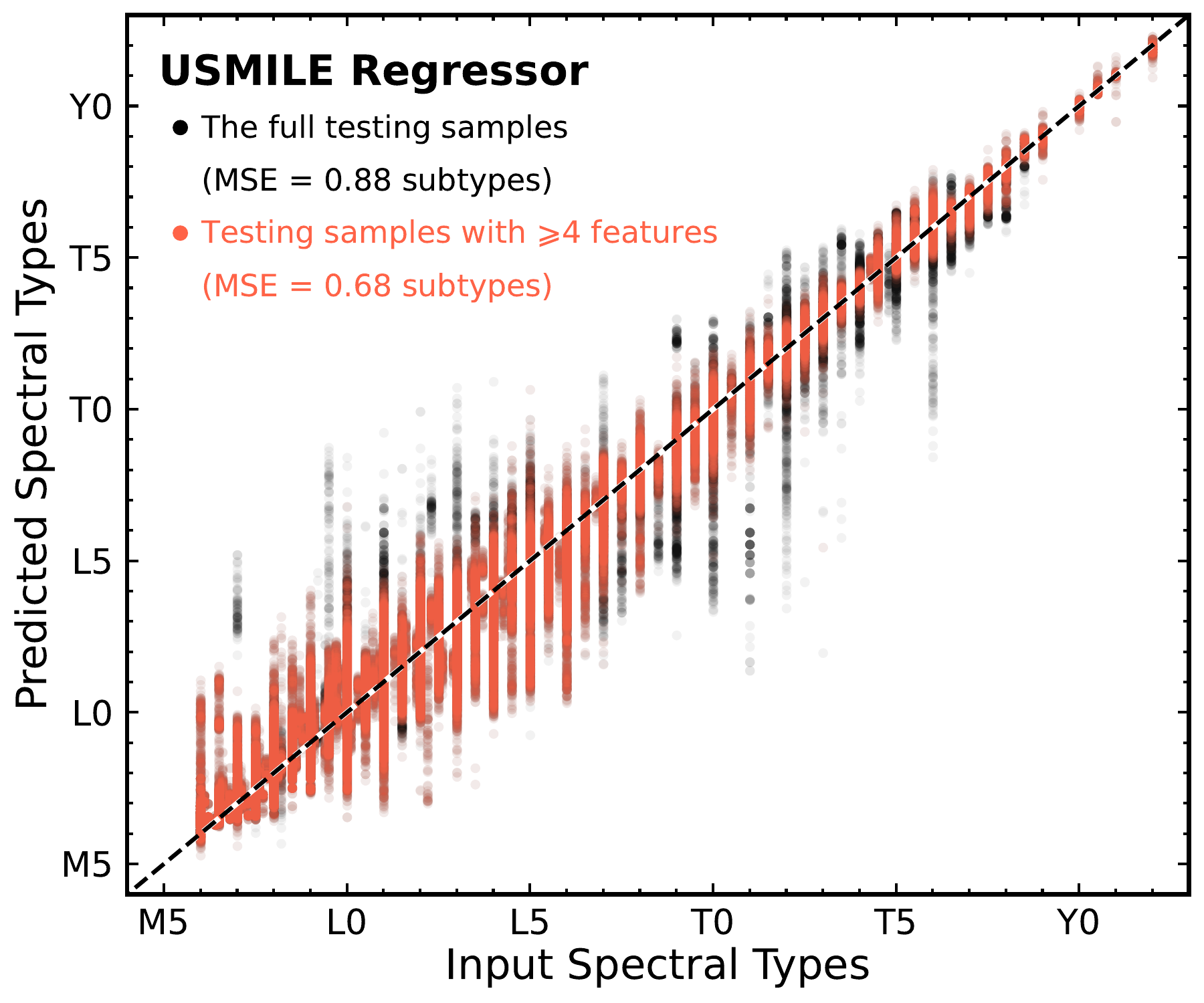}
\caption{Input versus predicted spectral types from the \texttt{USMILE regressor} for the full testing set (black). Objects with four or more available color features are highlighted in red. }
\label{fig:input_vs_predicted_spt}
\end{center}
\end{figure*}

\begin{figure}[t]
\begin{center}
\includegraphics[width=3.5in]{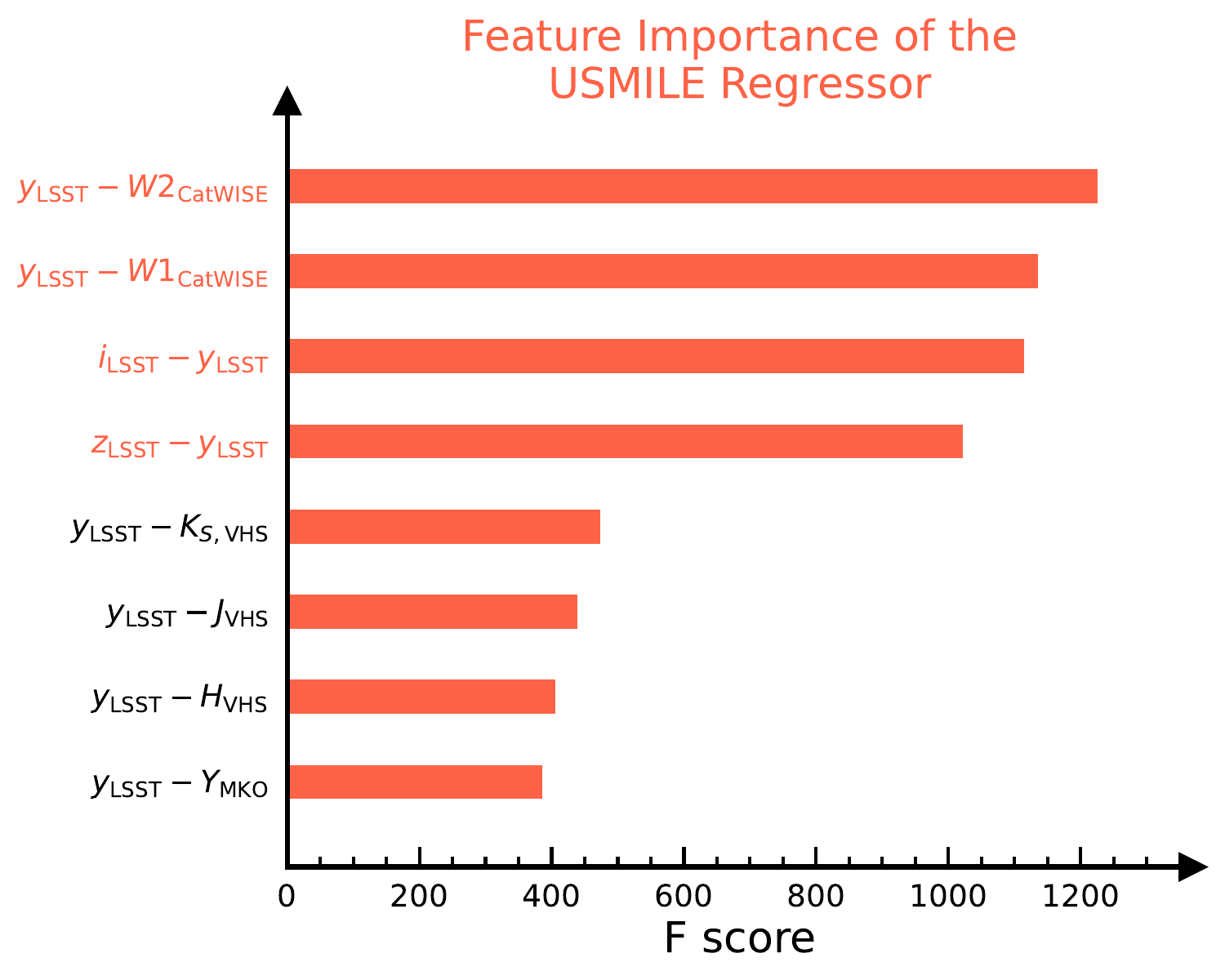}
\caption{Feature importance of the \texttt{USMILE regressor}, in the same format as Figure~\ref{fig:feature_important_classifier}  }
\label{fig:feature_important_regressor}
\end{center}
\end{figure}

\renewcommand{\arraystretch}{1.25} 
{ 
\begin{deluxetable*}{lccccc}
\setlength{\tabcolsep}{10pt} 
\tablecaption{Performance of the \texttt{USMILE Classifier} } \label{tab:performance} 
\tablehead{ \multicolumn{1}{l}{}   &   \multicolumn{2}{c}{\texttt{Baseline Classifier}}   &   \multicolumn{1}{c}{}   &   \multicolumn{2}{c}{\texttt{Customized Classifier}}    \\ 
\cline{2-3} \cline{5-6} 
\multicolumn{1}{l}{Metric}   &   \multicolumn{1}{c}{Training set}   &   \multicolumn{1}{c}{Testing set}   &   \multicolumn{1}{c}{}   &   \multicolumn{1}{c}{Training set}   &   \multicolumn{1}{c}{Testing set}    } 
\ 
\startdata 
ROC AUC & 0.9999999 & 0.9999996 & & 0.973 & 0.972 \\ 
True Negative & 1,007,114 & 252,152 & & 880,266 & 220,236 \\ 
False Positive & 147 & 87 & & 126,995 & 32,003 \\ 
False Negative & 44 & 36 & & 49,555 & 12,625 \\ 
True Positive & 1,025,895 & 256,025 & & 976,384 & 243,436 \\ 
Accuracy & 0.99991 & 0.99976 & & 0.913 & 0.912 \\ 
Precision & 0.99986 & 0.99966 & & 0.885 & 0.884 \\ 
Recall & 0.99996 & 0.99986 & & 0.952 & 0.951 \\ 
F1 Score & 0.9999 & 0.9998 & & 0.917 & 0.916 \\ 
\enddata 
\end{deluxetable*} 
}

\subsection{USMILE Regressor}
\label{subsec:regressor}

We further develop the \texttt{USMILE Regressor} to predict quantitative spectral types of ultracool dwarfs from color features. The model is implemented with XGBoost's gradient boosting decision tree regressor using a squared-error loss function, trained with the same number of trees, maximum tree depth, and learning rate as the classifier. Only positive samples from the labeled dataset are used, split into training and testing subsets in an 80:20 ratio. Unlike the \texttt{customized classifiers}, we do not mask feature values to mimic the completeness patterns of our classifier-selected candidates, since regression benefits from retaining a more complete set of features than is necessary for binary classification. We remain mindful, however, that \texttt{regressor} performance may depend on feature completeness, and we develop strategies to interpret the \texttt{regressor} outputs in the context of external spectroscopic validations Euclid Q1 (Section~\ref{sec:euclid}).

The \texttt{USMILE Regressor} achieves high accuracy, with mean squared errors of 0.86 and 0.88 subtypes for the training and testing subsets, respectively. The corresponding $R^2$ scores of 0.9714 and 0.9710 indicate that the model explains about $97\%$ of the variance in spectral types. Figure~\ref{fig:input_vs_predicted_spt} demonstrates the close agreement between input and predicted spectral types across the testing set. When restricting the testing set to objects with four or more available features, performance improves further, with an MSE of 0.68~subtypes and $R^{2}$ score of 0.973, along with fewer outliers across the L and T types. The four most influential color features for the \texttt{regressor} (Figure~\ref{fig:feature_important_regressor}) are $\ylsst-\wb$, $\ylsst-\wa$, $\ilsst-\ylsst$, and $\zlsst-\ylsst$, identical to those for the \texttt{customized classifier} (Figure~\ref{fig:feature_important_classifier}).

In the following sections, we apply the \texttt{USMILE customized classifiers} and \texttt{regressor} to candidates selected from LSST DP1, VHS, and CatWISE to identify new ultracool dwarf discoveries.

\section{Initial Search with LSST DP1, VHS, and CatWISE}
\label{sec:lsst_dp1}

We begin our search for ultracool dwarfs by mining LSST DP1 \citep{2025lsst.data....3N, 2025rubn.rept...31N}, released on June 30, 2025, using the Rubin Science Platform \citep[][]{2024ASPC..535..227O, juric2019lsstsp, DuboisFelsmann2017LDM542}. Among the seven fields included in LSST DP1, we focus on three that provide both $\zlsst$ and $\ylsst$ photometry, given that the $\zlsst-\ylsst$ color is one important feature for identifying ultracool dwarfs with the \texttt{USMILE classifier} and \texttt{regressor} (Section~\ref{sec:method}; Figures~\ref{fig:feature_important_classifier} and \ref{fig:feature_important_regressor}). These fields include the Extended Chandra Deep Field South (ECDFS), the EDF-S, and a low Galactic-latitude field (Rubin SV 95 -25). 

We query all LSST DP1 objects within a $1^{\circ}$ radius of each field center, retaining only non-blended sources classified as point-like in the $\ylsst$ band. This query is executed through a custom Python function built on the LSST Table Access Protocol (TAP) service, yielding 10,102 objects. For each source, we retrieve point-spread-function (PSF) magnitudes and their uncertainties in all available bands, and assign a good-quality flag if the object (1) is classified as a point-like source, (2) is not located on a CCD edge, (3) has unsaturated central pixels, and (4) is unaffected by known bad pixels. 

Next, we cross-match the LSST DP1 catalog with the VHS Data Release 5 \citep{2013Msngr.154...35M} and CatWISE2020 \citep{2021ApJS..253....8M}, using matching radii of $2''$ and $5''$, respectively; the larger radius for CatWISE accounts for its broader PSF. For VHS, we adopt photometry measured with a $2''$-diameter aperture and assign a good-quality flag in each band if the photometry is free from major catalog-assigned warnings. These warnings include cases where the sources (1) are close to saturation, (2) have potential photometric calibration issues, or (3) are located near frame boundaries or within underexposed regions. For CatWISE, we use profile-fitting photometry that accounts for proper motions and assign a good-quality flag when measurements are unaffected by known artifacts.

\begin{figure*}[t]
\begin{center}
\includegraphics[width=7in]{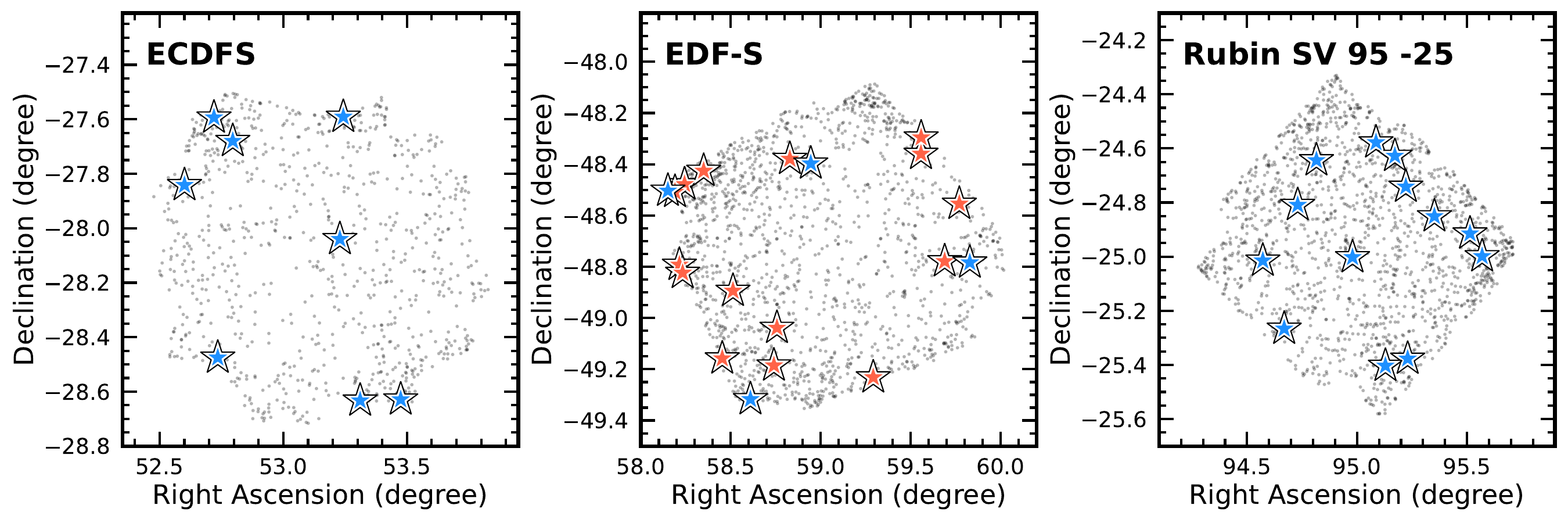}
\caption{Coordinates of the initial candidates (Section~\ref{sec:lsst_dp1}) selected from LSST DP1, VHS, and CatWISE (grey circles) within the Extended Chandra Deep Field South (ECDFS; left), Euclid Deep Field South (EDF-S; middle), and a Low Galactic Latitude Field (Rubin SV 95 -25; right). Orange stars mark the ultracool dwarf discoveries confirmed by Euclid spectroscopy, all located in the EDF-S field (Section~\ref{sec:euclid}). These discoveries span spectral types of M6--L2. Guided by the Euclid spectroscopic validation of the \texttt{USMILE} models, we further identify a set of high-quality M6--L9 candidates that lack high-S/N Euclid spectroscopy (Section~\ref{sec:final}), shown as blue stars.}
\label{fig:radec}
\end{center}
\end{figure*}

Anchored on the $\ylsst$ band, we compute eight colors for each object: $\ilsst - \ylsst$ and $\zlsst - \ylsst$, and $\ylsst - \yvhs$, $\ylsst - \jvhs$, $\ylsst - \hvhs$, $\ylsst - \kvhs$, $\ylsst - \wa$, and $\ylsst - \wb$ --- the same set of colors used to construct the labeled dataset (Section~\ref{sec:master_set}). Colors are calculated only when both contributing magnitudes have good-quality flags. Objects lacking any valid colors are removed. This is equivalent to requiring good-quality $\ylsst$ photometry plus at least one additional band with good-quality measurements --- a more flexible criterion than those adopted in many previous searches, which required detections in most or all available bands \citep[e.g.,][]{2019MNRAS.488.2263B, 2022A&A...666A.147M, 2022RNAAS...6...74G}. This filtering yields 4,053 initial candidates. The fraction of missing feature values is summarized in Table~\ref{tab:missing}, and their sky distribution is shown in Figure~\ref{fig:radec}.

\section{Applying the \texttt{USMILE} Models}
\label{sec:apply}

Our recommended procedure for confirming ultracool dwarf candidates in future surveys is to first apply the \texttt{USMILE customized classifiers} to identify high-quality photometric candidates and then use the \texttt{regressor} to estimate their spectral types. In this work, however, we apply both models to the full candidate list. This approach prepares us for external validation with Euclid Q1 spectra and for developing strategies to interpret model results when applied to other wide-field surveys (Section~\ref{sec:euclid}).

\begin{figure*}[t]
\begin{center}
\includegraphics[width=7.in]{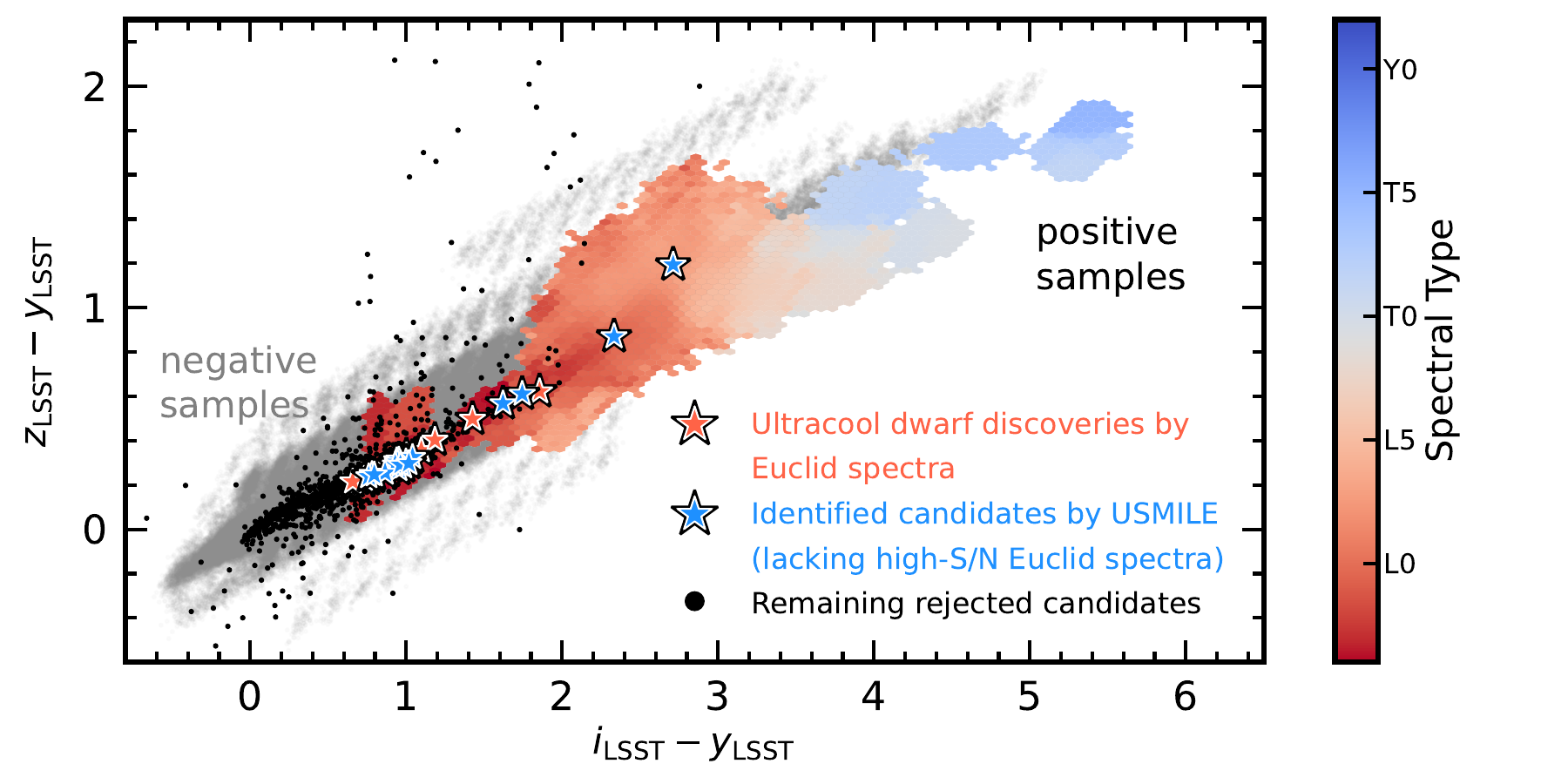}
\includegraphics[width=7.in]{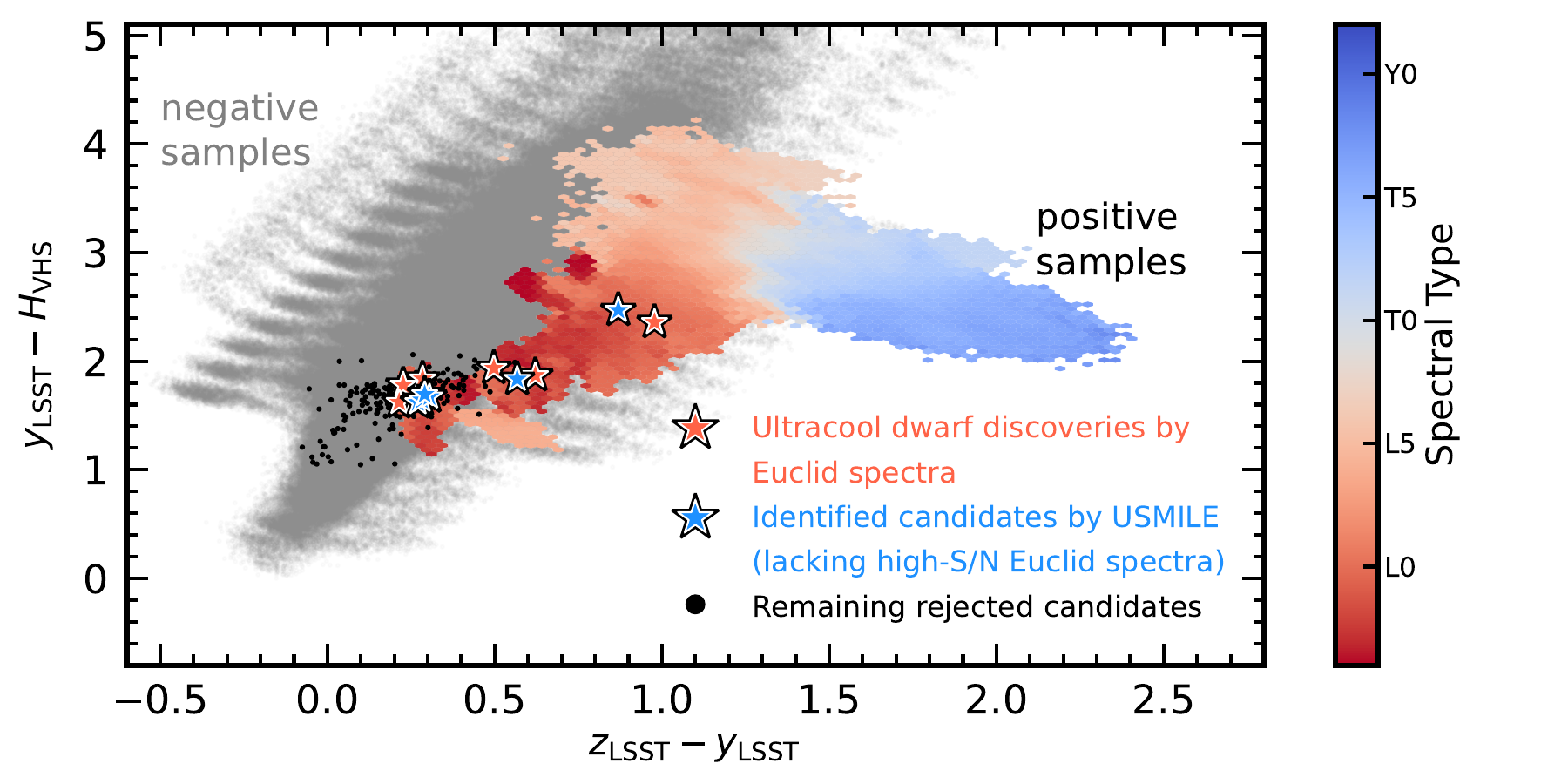}
\caption{{\it Top panel}: The $\ilsst-\ylsst$ and $\zlsst-\ylsst$ colors for the positive (colored) and negative (grey) samples in our labeled dataset, with positive samples color-coded by spectral types. The 15 ultracool dwarf discoveries confirmed with Euclid spectroscopy are shown as orange stars (Section~\ref{sec:euclid}). For objects lacking high-S/N Euclid spectra, we identify 25 high-quality candidates based on our \texttt{USMILE} models, shown as blue stars (Section~\ref{sec:final}). The remaining rejected candidates are shown as black circles. {\it Bottom panel}: The $\zlsst-\ylsst$ and $\ylsst - \hvhs$ colors for objects plotted in the top panel. }
\label{fig:iyzy_zyyh}
\end{center}
\end{figure*}

\begin{figure*}[t]
\begin{center}
\includegraphics[width=7.in]{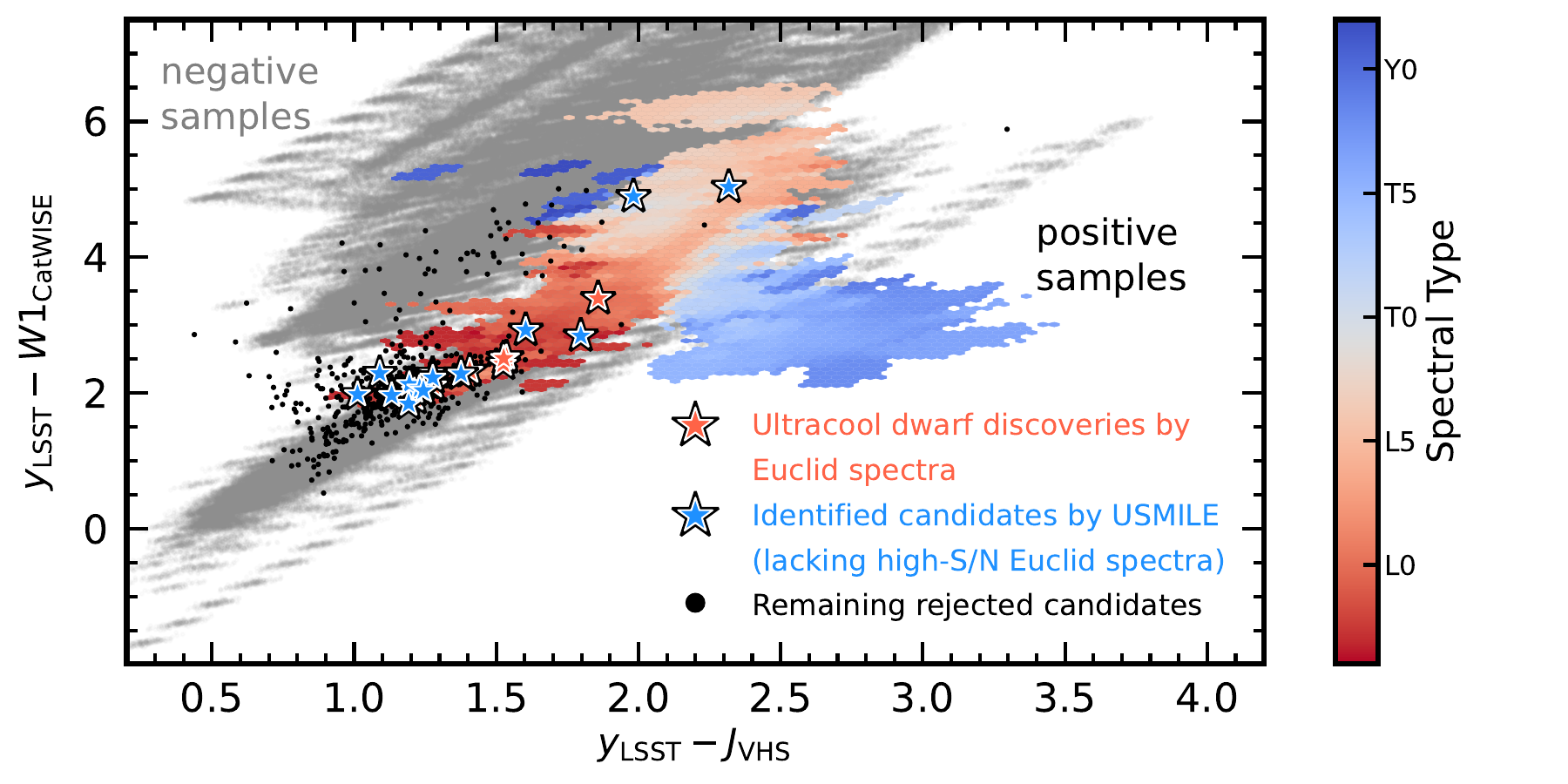}
\includegraphics[width=7.in]{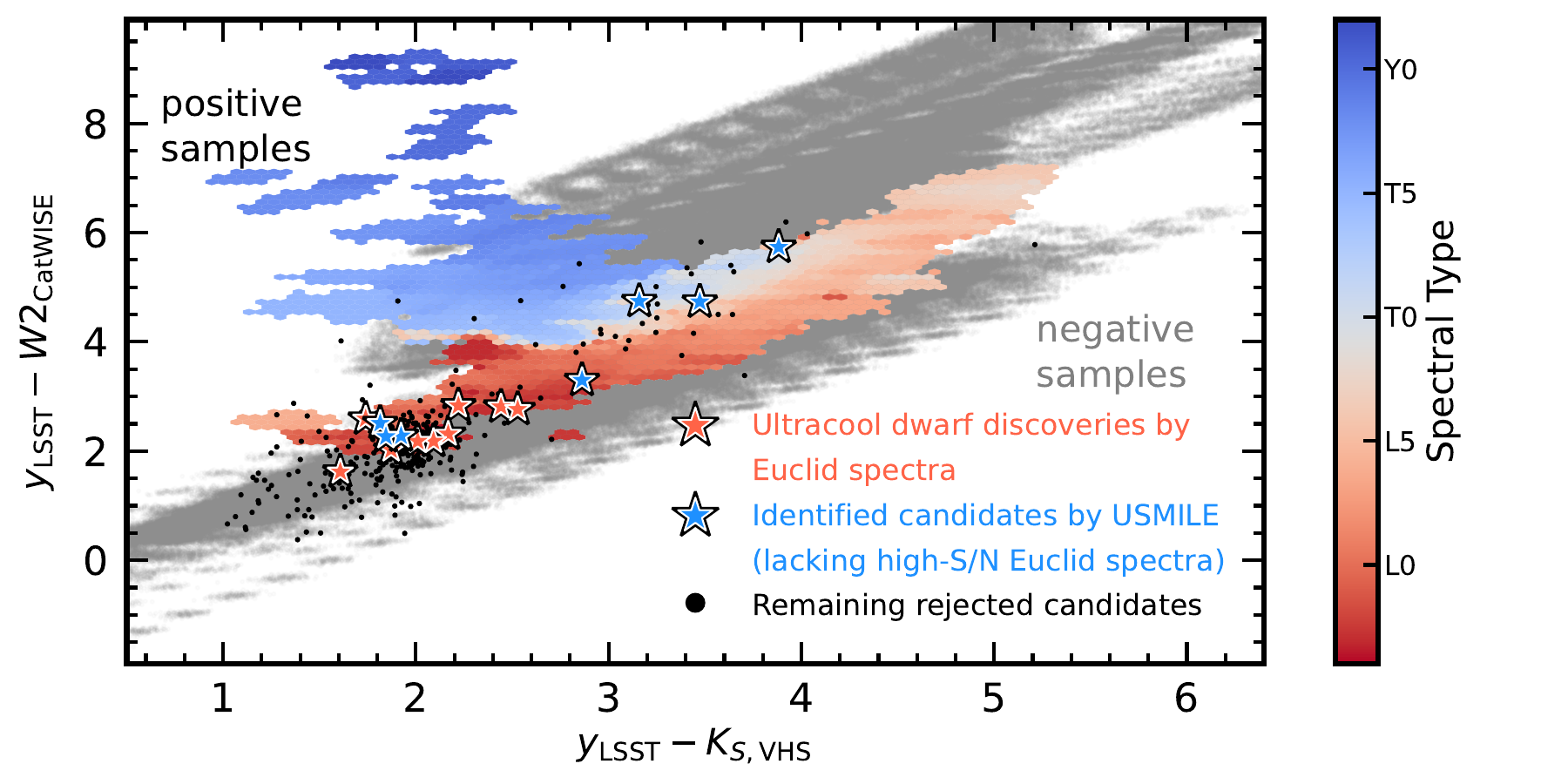}
\caption{The $\ylsst-\jvhs$ and $\ylsst-\wa$ colors (top), as well as the $\ylsst-\kvhs$ and $\ylsst-\wb$ colors (bottom), for the same type of objects plotted in Figure~\ref{fig:iyzy_zyyh}.}
\label{fig:yjyw1_ykyw2}
\end{center}
\end{figure*}

We apply all 400 \texttt{customized classifiers} to the candidate list and record each model's predicted binary label (ultracool dwarf vs. contaminant). Based on the aggregate voting across classifiers, we divide candidates into three categories:
\begin{enumerate}
\item[$\bullet$] \texttt{UCD-never}: never classified as an ultracool dwarf by any classifiers (0 out of 400); 2850 candidates ($70\%$).
\item[$\bullet$] \texttt{UCD-mixed}: classified as an ultracool dwarf by at least one but fewer than half of the classifiers (1--199 out of 400); 312 candidates ($8\%$).
\item[$\bullet$] \texttt{UCD-agreed}: classified as an ultracool dwarf by more than half of the classifiers (200--400 out of 400); 891 candidates ($32\%$).
\end{enumerate}
We then apply the \texttt{regressor} to all candidates in these three categories to predict spectral types, while also recording the number of available features for each candidate. The colors of all these candidates, compared with the positive and negative samples in our labeled dataset, are shown in Figures~\ref{fig:iyzy_zyyh} and \ref{fig:yjyw1_ykyw2}.

\begin{figure*}[t]
\begin{center}
\includegraphics[width=6.in]{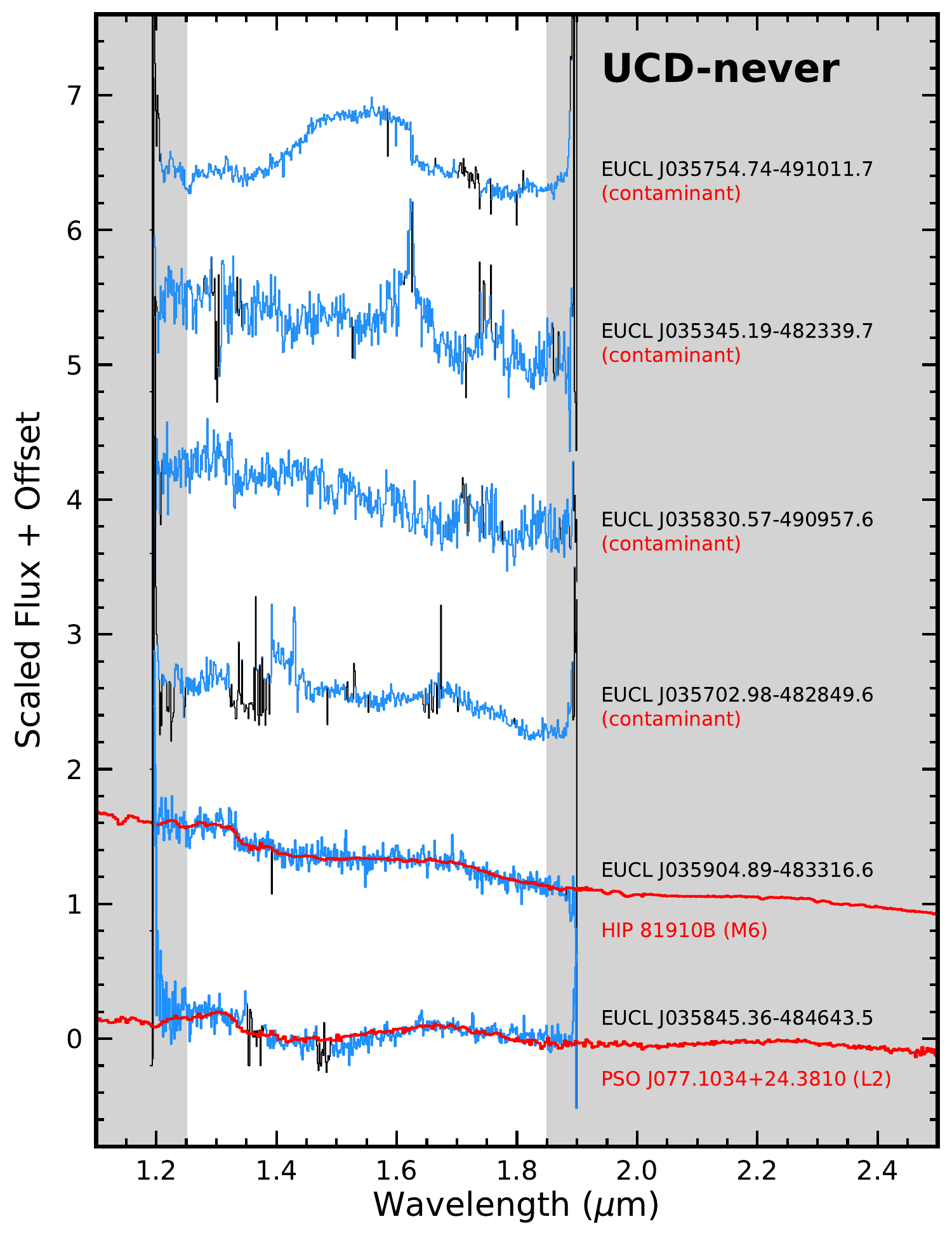}
\caption{Euclid NISP spectra for example candidates from the \texttt{UCD-never} category, which are classified as non-ultracool dwarfs by all 400 \texttt{customized classifiers}. Full NISP spectra are shown in black, with flux points from more than two dither positions (\texttt{NDITH}$>2$) highlighted in blue. The spectra of best-fit templates (red) are overlaid for candidates whose ultracool nature is confirmed by our visual inspection. The top four spectra are confirmed as contaminants based on our template fits and visual inspection, consistent with the classifiers' predictions. The second-to-last candidate is best matched to an M6 template, while the last is matched to an L2 template. The vertical white band marks the wavelength range (1.25--1.85~$\mu$m) used for template fitting analysis. }
\label{fig:never}
\end{center}
\end{figure*}

\begin{figure*}[t]
\begin{center}
\includegraphics[width=6.in]{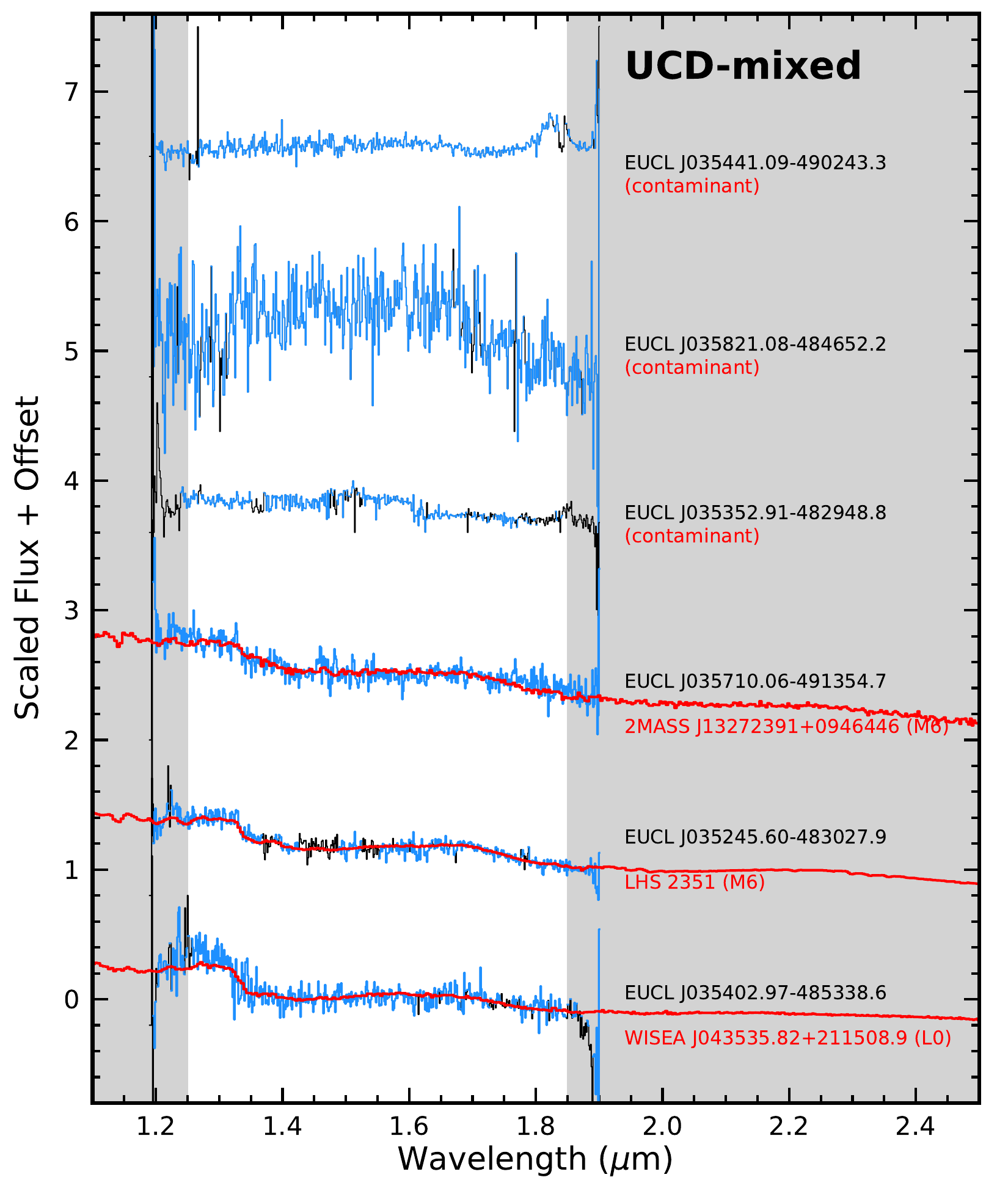}
\caption{Euclid NISP spectra for example candidates from the \texttt{UCD-mixed} category, which are classified as ultracool dwarfs by 1--199 out of 400 \texttt{customized classifiers}. The format follows that of Figure~\ref{fig:never}. }
\label{fig:mixed}
\end{center}
\end{figure*}

\begin{figure*}[t]
\begin{center}
\includegraphics[width=6.in]{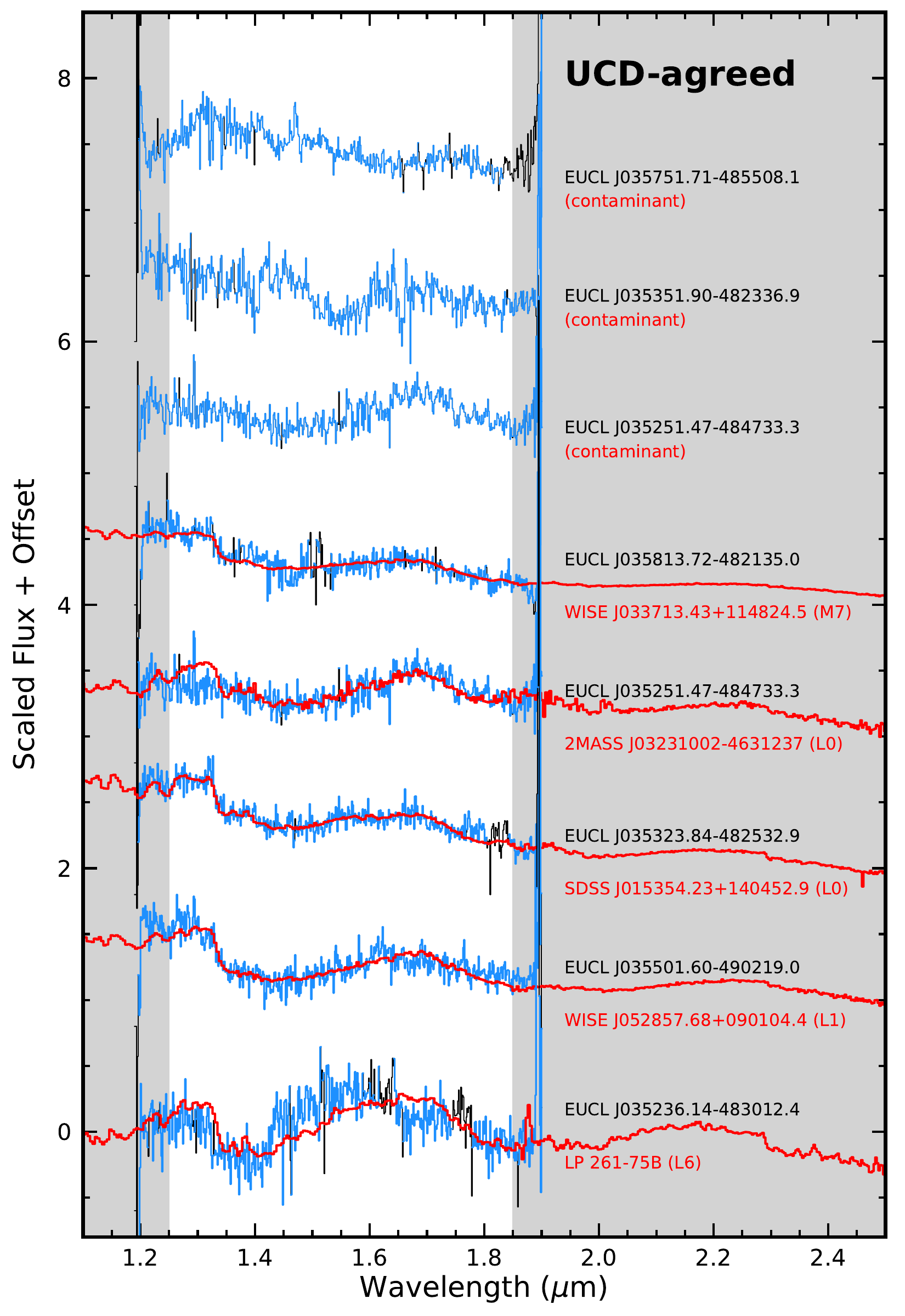}
\caption{Euclid NISP spectra for example candidates from the \texttt{UCD-agreed} category, with the format following that of Figure~\ref{fig:never}. The spectrum of the last candidate, EUCL~J035236.14$-$483012.4, only has a $J$-band S/N of 4.7. Although this S/N is lower than the threshold we adopt for performing template fitting analysis, we find its spectrum is similar to that of an L6 template. We keep this object as an ultracool dwarf candidate; high-quality spectroscopy is warranted to refine its spectral type and confirm its nature.  }
\label{fig:agreed}
\end{center}
\end{figure*}

\section{External Validation of \texttt{USMILE} models with Spectroscopy from Euclid Q1}
\label{sec:euclid}

One of the three LSST DP1 fields used in this work lies within the Euclid Deep Field South, offering a rare and valuable opportunity to perform large-scale external validation of our classifications using Euclid Q1 spectroscopy from the Near-Infrared Spectrometer and Photometer (NISP) \citep{2025A&A...697A...1E, 2025arXiv250315302E, 2025arXiv250315307E, 2025arXiv250315308E}. In the past, ultracool dwarf searches have relied on follow-up spectroscopy of only a handful to a few dozen candidates to validate the performance of their candidate identification methodologies \citep[e.g.,][]{2015MNRAS.450.2486C, 2016A&A...589A..49S, 2018ApJ...858...41Z, 2023MNRAS.522.1951D, 2024AJ....168..211B, 2024A&A...686A.171Z}. Such observations are important but necessarily resource-intensive, making it difficult to validate the methods on a large scale. Euclid now transforms this landscape by delivering spectra for hundreds of our classified candidates --- including both those identified as ultracool dwarfs and contaminants --- and, as we expand the searches across wider sky areas over the coming years, it will deliver millions of spectra. This capability enables a systematic test of the performance of our \texttt{USMILE} models. 

We cross-match all our EDF-S candidates against the Euclid \texttt{MER} and \texttt{spectra\_source} catalogs and download all available NISP spectra from ESA Datalabs \citep{Navarro2024}. Of the 1,555 EDF-S candidates, 1480 have Euclid counterparts and 865 have spectra, with a mean S/N of 7.2 per pixel in the $J$ band (1.2--1.3~$\mu$m) and a mode of 2.4. Restricting to spectra with $J$-band  S/N$\geqslant 5$ yields 291 sources suitable for testing the \texttt{USMILE} classifications and spectral type predictions. 

Our analysis is conducted in two steps. First, we compare each candidate's Euclid spectrum against a library of 1,263 M3--T9 templates, all with IRTF/SpeX spectra taken in prism mode ($R \sim 100$). Our template-fitting procedure follows \cite{2025AJ....169....9Z} and is performed via $\chi^{2}$ minimization over 1.25--1.85~$\mu$m, using all Euclid spectral fluxes in this range without quality cuts. Most spectral templates are selected from the UltracoolSheet \citep{ultracoolsheet} with additional M3--M6 templates collected from the literature. The complete list of references for these spectroscopic data is: \cite{2004ApJ...614L..73B, 2007ApJ...658..617B, 2007ApJ...659..655B}, 
\cite{2004AJ....127.2856B, 2006ApJ...639.1095B, 2006ApJ...637.1067B, 2007ApJ...658..557B, 2008ApJ...681..579B, 2008ApJ...674..451B, 2009ApJ...697..148B, 2010ApJ...710.1142B, 2010AJ....139.2448B, 2011AJ....141...70B, 2012ApJ...757..110B, 2013ApJ...772..129B, 2015AJ....149..104B}, 
\cite{2004ApJ...604L..61C}, 
\cite{2004PASP..116..362C}, 
\cite{2006AJ....131.1007B, 2006ApJ...645.1485B}, 
\cite{2006AJ....131.2722C}, 
\cite{2006AJ....132..891R, 2006ApJ...639.1114R}, 
\cite{2006AJ....132.2074M}, 
\cite{2006ApJ...639.1120K, 2008ApJ...689.1295K, 2010ApJS..190..100K, 2011ApJS..197...19K, 2014ApJ...783..122K}, 
\cite{2007AJ....133.2320S}, 
\cite{2007AJ....134..411M}, 
\cite{2007ApJ...669L..97L, 2007AJ....134.1162L, 2008ApJ...686..528L}, 
\cite{2007ApJ...654..570L, 2012ApJ...760..152L}, 
\cite{2007ApJ...655..522L}, 
\cite{2008ApJ...676.1281M}, 
\cite{2009AJ....137..304S}, 
\cite{2009ApJ...706.1114B, 2012ApJ...753..142B, 2013ApJ...774...55B, 2014ApJ...784...65B}, 
\cite{2010AJ....139..176F, 2011AJ....141...71F, 2016ApJS..225...10F}, 
\cite{2010AJ....139.1045S}, 
\cite{2010ApJ...715..561A}, 
\cite{2011AJ....141....7D}, 
\cite{2011AJ....142...77D, 2012ApJ...755...94D, 2012ApJ...757..100D, 2014ApJ...792..119D, 2017MNRAS.467.1126D}, 
\cite{2011AJ....142..171G, 2011ApJ...736L..34G, 2012AJ....144...94G}, 
\cite{2011ApJ...726...30M}, 
\cite{2011ApJ...732...56G},
\cite{2012ApJS..201...19D}, 
\cite{2013ApJ...772...79A}, 
\cite{2013ApJ...776..126C}, 
\cite{2013ApJ...777...84B, 2015ApJ...814..118B, 2017ApJ...843L...4B, 2017ApJ...837...95B}, 
\cite{2013ApJ...777L..20L, 2016ApJ...833...96L}, 
\cite{2013ApJS..205....6M}, 
\cite{2013PASP..125..809T}, 
\cite{2014AJ....147...34S, 2015ApJ...804...92S}, 
\cite{2014ApJ...785L..14G, 2015ApJS..219...33G}, 
\cite{2014ApJ...787..126L}, 
\cite{2014ApJ...794..143B}, 
\cite{2016AJ....151...46A}, 
\cite{2016ApJ...821..120A}, 
\cite{2016Natur.533..221G}, 
\cite{2016PhDT.......189A}, 
\cite{2017AJ....154..112K}, 
\cite{2017ASInC..14....7B}, 
\cite{2021ApJ...911....7Z, 2021ApJ...921...95Z}, 
\cite{2023ApJ...959...63S}, 
and \cite{2024ApJ...961..121H}.

Second, we visually inspect template fits to assess whether the spectra have sufficient quality for reliable template fitting, whether they are well matched to the best-fitting templates, and whether they exhibit hallmark ultracool dwarf features, such as the broad H$_{2}$O absorption feature between the $J$ and $H$ bands (1.3--1.5~$\mu$m). During inspection, we highlight spectral fluxes obtained from more than two dithers \citep[\texttt{NDITH}$>2$;][]{2025arXiv250322442D} and require that these good-quality pixels span at least half of the wavelength range or, at least, cover the peak of either the $J$ or $H$ band.

A total of 171 spectra meet these quality requirements for template-fitting: 129 from \texttt{UCD-never}, 26 from \texttt{UCD-mixed}, and 16 from \texttt{UCD-agreed}. Their properties are discussed below.

\subsection{129 \texttt{UCD-never} Candidates}
\label{subsec:never}
By definition, these candidates were classified as contaminants by all 400 \texttt{customized classifiers} (Section~\ref{sec:apply}). Template fitting and visual inspection confirm 122 as contaminants, fully consistent with the classifiers' predictions. 

The remaining 7 objects have color features located near the boundaries between the positive and negative samples in the labeled dataset (Figures~\ref{fig:iyzy_zyyh} and \ref{fig:yjyw1_ykyw2}). Six are best matched to M6--M8 templates, and their five best-fitting templates span a 1--5 subtype range. Five of these have $\geqslant 4$ color features, with \texttt{regressor} predictions of M6--M9, consistent with template-based spectral types; the other one has only three features, with the \texttt{regressor}-predicted spectral type of L3.1 and template-based type of M6. 

The final object, LSST-DP1-O-592913775581988519 (or EUCL~J035845.36$-$484643.5), is best matched to an L2 template, with its five best-fitting templates spanning M8.5--L2. This object has seven color features, and the \texttt{regressor} predicts a spectral type of M8.2, broadly consistent with the template fits, given its modest S/N of 8.9 in the $J$ band. 

Overall, these results show that the false negatives of our models are rare, and arise mainly near the M6 boundary, where classification is inherently ambiguous. Moreover, the \texttt{regressor} predictions are consistent with template fits, with scatter driven by low S/Ns of Euclid spectra or too few ($<4$) available features of these candidates. Example Euclid spectra of \texttt{UCD-never} candidates are shown in Figure~\ref{fig:never}.

\subsection{26 \texttt{UCD-mixed} Candidates}
\label{subsec:mixed}

These candidates were classified as ultracool dwarfs by only 1--199 out of 400 \texttt{customized classifiers} (Section~\ref{sec:apply}). Our template fitting confirms 22 as contaminants, consistent with the majority vote of these \texttt{classifiers}. 

The remaining 4 objects have colors located near the boundaries between the positive and negative samples. Three of these are best matched to M6--M6.5 templates, with \texttt{regressor} predictions of M6.5--M6.9, consistent with the template-based results. The last object, LSST-DP1-O-592913363265129113 (EUCL~J035402.97$-$485338.6), has four available color features and is best matched to an L0 template, with its five best-fitting templates spanning M9--L0 types. However, the \texttt{regressor} predicts M6.6. This discrepancy likely reflects an uncertainty in the template-based spectral type, given the moderate S/N of 7.5 for its Euclid spectrum, and the intrinsic scatter of \texttt{regressor}'s predictions.

Thus, \texttt{UCD-mixed} candidates are dominated by contaminants with few borderline M6 objects. Discrepancies between template- and \texttt{regressor}-based spectral types occur primarily when Euclid spectra are noisy. Example Euclid spectra of \texttt{UCD-mixed} candidates are shown in Figure~\ref{fig:mixed}.

\subsection{16 \texttt{UCD-agreed} Candidates}
\label{subsec:agreed}

These candidates were identified as ultracool dwarfs by 200--400 out of 400 \texttt{customized classifiers}. Of these, 12 candidates are confirmed contaminants, all with $\leqslant 4$ available color features (six with 2 features, one with 3 features, and five with 4 features). The remaining 4 candidates are confirmed ultracool dwarfs, all with $\geqslant 4$ available color features (one each with 4, 5, 6, and 7 features.

Among the 12 contaminants, most objects have only $\ylsst-\wa$ and $\ylsst-\wb$, or $\ilsst-\ylsst$ and $\zlsst-\ylsst$ colors, which are highly correlated (with Pearson coefficients of 0.97; Figure~\ref{fig:classifier_corr}). Their limited information content explains why  \texttt{classifiers} misidentify them. This result highlights that \texttt{UCD-agreed} predictions based on fewer than four features --- especially correlated ones --- should be treated with caution.

Among the 4 ultracool dwarfs, their template-based spectral types span M7--L1. For one object with $J$-band S/N$>15$, \texttt{regressor} and template fits yield consistent spectral types within 0.5 subtypes. For the remaining three objects with lower S/Ns (8--9), the \texttt{regressor} predicts L5.5, M6.6, and M6.7, while their five best-fitting templates span M8.5--L1, L0--L5, and M7--M9.5, respectively. These discrepancies are likely due to the moderate S/Ns of their Euclid spectra and the intrinsic scatter in the \texttt{regressor}'s predictions (Figure~\ref{fig:input_vs_predicted_spt}). Example Euclid spectra of \texttt{UCD-agreed} candidates are presented in Figure~\ref{fig:agreed}. 

One additional object, LSST-DP1-O-592908071865418214 (or EUCL~J035236.14$-$483012.4), with a $J$-band spectral S/N of 4.7, falls just below our adopted S/N threshold of $\geqslant 5$ for template fitting. Nevertheless, its spectrum is similar to that of an L6 template, showing the hallmark H$_{2}$O absorption feature, though with an anomalous $H$-band peak. The \texttt{regressor} predicts a spectral type of L3.6, albeit from only two features ($\ylsst-\wa$ and $\ylsst-\wb$). We consider this spectral type tentative. Spectroscopic follow-up of this object is useful to refine its ultracool nature and spectral type.

\subsection{Strategies for Interpreting the \texttt{USMILE} Model Results }
\label{subsec:strategies}

The performance of \texttt{USMILE} models, as validated by Euclid Q1, can be summarized as:
\begin{itemize}
\item[$\bullet$] \texttt{UCD-never} candidates are overwhelmingly contaminants, with the few exceptions with M6 types.  
\item[$\bullet$] \texttt{UCD-mixed} candidates are mostly contaminants, with the few exceptions with M6 types.
\item[$\bullet$] \texttt{UCD-agreed} candidates are generally ultracool dwarfs when more than 4 color features are available, and false positives arise when feature counts are below 4. Candidates with exactly 4 features show mixed outcomes, including both genuine ultracool dwarfs and contaminants. 
\end{itemize}
In summary, Euclid Q1 spectroscopy offers a rare and large-scale external validation of our methods, confirming that the \texttt{USMILE classifiers} and \texttt{regressor} deliver accurate predictions in practice, while clarifying the regimes where these predictions are most reliable. Beyond the strong performance with training and testing sets, this Euclid-based validation again demonstrates that the \texttt{USMILE} models are robust and interpretable, and ready to be applied to future surveys.

\section{Ultracool Candidates and Discoveries}
\label{sec:final}

Table~\ref{tab:summary} lists the 15 ultracool dwarf discoveries confirmed through Euclid spectroscopy (Section~\ref{sec:euclid}). These objects have spectral types from M6 to L2, based on their best-fitting templates. We also include 1 preliminary L6 dwarf, LSST-DP1-O-592908071865418214 (or EUCL~J035236.14$-$483012.4), discussed in Section~\ref{subsec:agreed}, whose ultracool nature is tentative and requires spectroscopic follow-up.

All these confirmed discoveries are identified as point sources in the Euclid \texttt{MER} catalog, with probabilities ranging from 0.9 to 1.0. These values are consistent with their point-like source classifications by LSST DP1.

Beyond the spectroscopically confirmed discoveries, we compile an additional list of 24 high-quality ultracool dwarf candidates among those lacking high-S/N  Euclid spectra. These candidates have the \texttt{regressor}-based spectral types of M6--L9 and are selected using the validated strategies described in Section~\ref{subsec:strategies}:
\begin{enumerate}
\item[$\bullet$] They are classified as ultracool dwarfs by at least half of the \texttt{customized classifiers}. In other words, they fall in the \texttt{UCD-agreed} category.
\item[$\bullet$] They have $\geqslant 4$ available color features.
\end{enumerate}
These candidates are summarized in Table~\ref{tab:summary}.

Finally, we note that one candidate, LSST-DP1-O-591818593281245219 (or EUCL~J035426.27-491902.2), has a near-zero point-source probability based on Euclid but is classified as a point source in LSST DP1. Given its relatively late spectral type of L8.3 as predicted by the \texttt{regressor}, we retain it in our candidate list, though spectroscopic follow-up is required to confirm its nature.

Together, we obtain 15 Euclid-confirmed M6--L2 discoveries and 25 (i.e., 1 preliminary L6 candidate with a low-S/N Euclid spectrum and 24 photometric candidates) high-priority M6--L9 candidates, thanks to the power of combining LSST photometry with Euclid spectroscopy.

\section{Summary}
\label{sec:summary}

We have presented the Ultracool dwarf Science with MachIne LEarning (USMILE) program and developed scalable, tree-based machine-learning tools for identifying ultracool dwarfs from wide-field sky surveys, including Rubin Observatory LSST DP1, VHS, and CatWISE. Our key results are summarized below. 

\begin{enumerate}
\item[1.] We assembled a comprehensive labeled dataset containing known ultracool dwarfs (spectral type $\geqslant$M6) and representative contaminants, including reddened early-type stars and quasars. By compiling or synthesizing LSST, VHS, and CatWISE photometry and applying data augmentation to propagate photometric uncertainties, we built a balanced labeled dataset of over 2 million objects. This dataset provides the foundation for robust model training and testing.

\item[2.] Using $\ylsst$ as the anchor band, we computed eight colors that are distance-independent: $\ilsst - \ylsst$, $\zlsst - \ylsst$, $\ylsst - \yvhs$, $\ylsst - \jvhs$, $\ylsst - \hvhs$, $\ylsst - \kvhs$, $\ylsst - \wa$, and $\ylsst - \wb$. These colors are treated as input features for our machine-learning models.

\item[3.] Two complementary tree-based models were developed with XGBoost: the \texttt{USMILE classifier}, a gradient-boosted decision tree classifier that distinguishes ultracool dwarfs from contaminants, and the \texttt{regressor}, which predicts quantitative spectral types. Both models are scalable to the massive data volumes of modern sky surveys. Another key strength is their ability to natively handle missing feature values, avoiding the need for imputation and its potential biases. This enables classification of candidates with incomplete photometry, which are common in wide-field searches, whereas earlier studies employing methods such as PCA, SVM, KNN, RF, and ANN generally had to restrict candidates to those detected in all relevant bands or rely on imputed values.

\item[4.] To address the incomplete color features of our initial candidates queried from LSST, VHS, and CatWISE, we trained 400 \texttt{customized classifiers} on ``downgraded'' labeled datasets with missing-value patterns matching those of the candidates. These \texttt{classifiers} achieve strong performance on both training and testing sets, with an ROC AUC of 0.97 and an F1 score of 0.92. Learning curves demonstrate that these models generalize effectively with well-balanced bias--variance trade-offs. The \texttt{regressor} yields a mean squared error of 0.88 subtypes on the testing set. For both the \texttt{classifier} and \texttt{regressor} models, the most influential features are $\ylsst-\wa$, $\ilsst-\ylsst$, $\zlsst-\ylsst$, and $\ylsst-\wb$ colors. 

\item[5.] We carried out the first systematic ultracool dwarf search using Rubin Observatory LSST DP1, focusing on three fields with both $\zlsst$ and $\ylsst$ coverage and cross-matched them with VHS and CatWISE. This yielded 4,053 initial candidates, to which we further applied the \texttt{customized classifiers} and \texttt{regressor} to assign classifications (ultracool dwarf vs. contaminant) and predicted spectral types.

\item[6.] A major advance of this work is the large-scale external validation enabled by Euclid Q1 spectroscopy. In the Euclid Deep Field South, 291 of our candidates --- including both those identified as ultracool dwarfs and those identified as contaminants by the \texttt{USMILE classifiers} --- have available Euclid NISP spectra with $J$-band S/N$\geqslant 5$. Whereas previous ultracool dwarf searches could spectroscopically confirm only a handful to a few dozen objects, Euclid delivered spectra for hundreds of our classified candidates in a single release. This rare and timely validation verified the strong performance of our \texttt{USMILE} models and confirmed 15 ultracool dwarf discoveries with types of M6--L2, plus one tentative L6 dwarf with a low-S/N Euclid spectrum.

\item[7.] Euclid validation also revealed the conditions under which our model predictions are most reliable: candidates are very likely ultracool dwarfs if they (1) are classified as such by at least half of the \texttt{customized classifiers}, and (2) have four or more (out of eight) available color features. Guided by these criteria, we compiled an additional list of 24 high-quality candidates that have no high-S/N Euclid spectra. Considering the aforementioned tentative L6 dwarf candidate with low-S/N Euclid spectrum, we thus derive a list of 25 candidates with the \texttt{regressor}-based spectral types of M6--L9. These constitute the most promising targets for future spectroscopic follow-up.
\end{enumerate}

Our study presents the first ultracool dwarf search using LSST DP1, uniquely combined with large-scale external validation from Euclid Q1 spectrocopy. These analyses confirm the strong performance of \texttt{USMILE} models and demonstrate how scalable machine-learning techniques can achieve accurate and efficient classification of ultracool dwarfs in the data-rich era of wide-field sky surveys. 

The synergy between LSST and Euclid is particularly powerful: LSST provides deep, precise optical photometry and astrometry, while Euclid adds complementary near-infrared photometry and spectroscopy, enabling rapid validation at a scale previously unattainable. VHS and CatWISE further enrich the number of available color features, improving the accuracy of both classification and spectral type estimates. 

Looking ahead, forthcoming LSST and Euclid data releases will provide greater depth and sky coverage, uncovering new ultracool dwarfs with even later spectral types than those identified in this work. These datasets will also allow  \texttt{USMILE} models to be applied to increasingly larger samples, providing a robust, data-driven framework to expand the census of stellar, substellar, and planetary-mass populations in the solar neighborhood.

\begin{acknowledgments}
We thank the anonymous referee for providing suggestions that improved the manuscript. This study is based upon work supported in part by the National Science Foundation through Cooperative Agreements AST-1258333 and AST-2241526 and Cooperative Support Agreements AST-1202910 and 2211468 managed by the Association of Universities for Research in Astronomy (AURA), and the Department of Energy under Contract No. DE-AC02-76SF00515 with the SLAC National Accelerator Laboratory managed by Stanford University. Additional Rubin Observatory funding comes from private donations, grants to universities, and in-kind support from LSST-DA Institutional Members. This work has made use of the Euclid Q1 data from the {\it Euclid} mission of the European Space Agency (ESA), 2025, \url{[https://doi.org/10.57780/esa-2853f3b|https://doi.org/10.57780/esa-2853f3b]}. This research has made use of the SVO Filter Profile Service ``Carlos Rodrigo'', funded by MCIN/AEI/10.13039/501100011033/ through grant PID2023-146210NB-I00. We made use of the data from the SpeX Prism Library service developed by the Spanish Virtual Observatory in the framework of the IAU Comission G5 Working Group : Spectral Stellar Libraries. This work has benefitted from The UltracoolSheet,\footnote{\url{http://bit.ly/UltracoolSheet}} maintained by Will Best, Trent Dupuy, Michael Liu, Aniket Sanghi, Rob Siverd, and Zhoujian Zhang, and developed from compilations by \cite{2012ApJS..201...19D}, \cite{2013Sci...341.1492D}, \cite{2016ApJ...833...96L}, \cite{2018ApJS..234....1B}, \cite{2021AJ....161...42B}, \cite{2023ApJ...959...63S}, and \cite{2023AJ....166..103S}. This research makes use of ESA Datalabs (datalabs.esa.int), an initiative by ESA's Data Science and Archives Division in the Science and Operations Department, Directorate of Science.

This paper and the USMILE program are dedicated to our daughter, Allison. The model introduced here, \texttt{USMILE Avocado}, is named after Allison's very first solid food. As parents, we hope U~SMILE every day $\smiley$!

\end{acknowledgments}

\vspace{5mm}
\facilities{Rubin:Simonyi (LSSTComCam), Euclid (NISP)}
\software{\texttt{XGBoost} \citep{2016arXiv160302754C}, \texttt{scikit-learn} \citep{scikit-learn}, \texttt{STILTS} \citep{2006ASPC..351..666T}, \texttt{TOPCAT} \citep{2005ASPC..347...29T}, \texttt{astropy} \citep{2013A&A...558A..33A, 2018AJ....156..123A, 2022ApJ...935..167A}, \texttt{ipython} \citep{PER-GRA:2007}, \texttt{numpy} \citep{numpy}, \texttt{scipy} \citep{scipy}, \texttt{matplotlib} \citep{Hunter:2007}.}

\clearpage

{ 
\renewcommand{\arraystretch}{1.25} 
\begin{longrotatetable}
{ 
\begin{deluxetable*}{llcccccccl}
\setlength{\tabcolsep}{4pt} 
\tablecaption{Ultracool Dwarf Discoveries and Candidates} \label{tab:summary} 
\tablehead{ \multicolumn{1}{l}{}   &   \multicolumn{1}{l}{}   &   \multicolumn{1}{l}{}   &   \multicolumn{1}{l}{}   &   \multicolumn{3}{c}{Example Photometry\tablenotemark{\scriptsize a}}   &   \multicolumn{1}{c}{}   &   \multicolumn{1}{c}{}   &   \multicolumn{1}{l}{}    \\ 
\cline{5-7} 
\multicolumn{1}{l}{LSST DP1 Object ID}   &   \multicolumn{1}{l}{Euclid ID}   &   \multicolumn{1}{c}{RA (LSST DP1)}   &   \multicolumn{1}{c}{DEC (LSST DP1)}   &   \multicolumn{1}{c}{$y_{\rm LSST}$}   &   \multicolumn{1}{c}{$J_{\rm VHS}$}   &   \multicolumn{1}{c}{$W1_{\rm CatWISE}$}   &   \multicolumn{1}{c}{$N_{\rm feature}$}   &   \multicolumn{1}{c}{SpT\tablenotemark{\scriptsize b}}   &   \multicolumn{1}{l}{Label}    \\ 
\multicolumn{1}{l}{}   &   \multicolumn{1}{l}{}   &   \multicolumn{1}{c}{(hh:mm:ss.ss)}   &   \multicolumn{1}{c}{(dd:mm:ss.s)}   &   \multicolumn{1}{c}{(AB mag)}   &   \multicolumn{1}{c}{(Vega mag)}   &   \multicolumn{1}{c}{(Vega mag)}   &   \multicolumn{1}{c}{}   &   \multicolumn{1}{c}{}   &   \multicolumn{1}{l}{}    } 
\ 
\startdata 
\hline 
\multicolumn{10}{c}{Ultracool Dwarf Discoveries (M6--L2)} \\ 
\hline 
592914187898847313 & EUCL J035251.47-484733.3 & 03:52:51.46 & $-$48:47:33.4 & $19.917 \pm 0.019$ & -- & $17.900 \pm 0.075$ & 4 & L0.0 & UCD-agreed \\ 
592915493568906539 & EUCL J035323.84-482532.9 & 03:53:23.84 & $-$48:25:33.0 & $19.444 \pm 0.015$ & $17.920 \pm 0.039$ & $17.016 \pm 0.039$ & 6 & L0.0 & UCD-agreed \\ 
592912676070361415 & EUCL J035501.60-490219.0 & 03:55:01.60 & $-$49:02:19.1 & $20.637 \pm 0.019$ & $18.779 \pm 0.056$ & $17.245 \pm 0.044$ & 5 & L1.0 & UCD-agreed \\ 
592915149971524359 & EUCL J035813.72-482135.0 & 03:58:13.72 & $-$48:21:35.0 & $19.672 \pm 0.016$ & $18.146 \pm 0.036$ & $17.160 \pm 0.042$ & 7 & M7.0 & UCD-agreed \\ 
592915562288382147 & EUCL J035245.60-483027.9 & 03:52:45.59 & $-$48:30:27.9 & $19.664 \pm 0.037$ & $18.129 \pm 0.048$ & $17.142 \pm 0.040$ & 5 & M6.5 & UCD-mixed \\ 
592914187898847267 & EUCL J035255.78-484926.6 & 03:52:55.78 & $-$48:49:26.6 & $19.888 \pm 0.018$ & -- & $17.919 \pm 0.077$ & 4 & M6.0 & UCD-mixed \\ 
592913363265129113 & EUCL J035402.97-485338.6 & 03:54:02.97 & $-$48:53:38.6 & $19.713 \pm 0.008$ & -- & $17.279 \pm 0.047$ & 4 & L0.0 & UCD-mixed \\ 
591818387122816088 & EUCL J035710.06-491354.7 & 03:57:10.06 & $-$49:13:54.7 & $19.684 \pm 0.018$ & -- & $17.640 \pm 0.060$ & 4 & M6.0 & UCD-mixed \\ 
592915562288382171 & EUCL J035258.35-482841.6 & 03:52:58.35 & $-$48:28:41.6 & $19.465 \pm 0.030$ & $18.252 \pm 0.052$ & $17.541 \pm 0.056$ & 5 & M6.0 & UCD-never \\ 
591819280476014638 & EUCL J035348.63-490929.3 & 03:53:48.63 & $-$49:09:29.3 & $19.208 \pm 0.014$ & -- & $17.082 \pm 0.040$ & 3 & M6.0 & UCD-never \\ 
591818524561771172 & EUCL J035457.57-491108.3 & 03:54:57.57 & $-$49:11:08.3 & $19.251 \pm 0.007$ & $18.010 \pm 0.028$ & $17.226 \pm 0.043$ & 7 & M8.0 & UCD-never \\ 
592915356129955036 & EUCL J035518.82-482245.7 & 03:55:18.81 & $-$48:22:45.7 & $18.990 \pm 0.007$ & $17.714 \pm 0.022$ & $16.712 \pm 0.030$ & 7 & M6.0 & UCD-never \\ 
592915837166289513 & EUCL J035813.97-481745.2 & 03:58:13.97 & $-$48:17:45.2 & $18.565 \pm 0.012$ & $17.286 \pm 0.017$ & $16.455 \pm 0.027$ & 7 & M6.0 & UCD-never \\ 
592913775581988519 & EUCL J035845.36-484643.5 & 03:58:45.36 & $-$48:46:43.5 & $20.036 \pm 0.015$ & $18.828 \pm 0.066$ & $17.935 \pm 0.077$ & 7 & L2.0 & UCD-never \\ 
592914462776757157 & EUCL J035904.89-483316.6 & 03:59:04.89 & $-$48:33:16.6 & $19.454 \pm 0.013$ & $18.048 \pm 0.032$ & $17.129 \pm 0.039$ & 7 & M6.0 & UCD-never \\ 
\hline 
\multicolumn{10}{c}{Ultracool Dwarf Candidates (M6--L9)} \\ 
\hline 
611255965995499968 & -- & 03:30:24.15 & $-$27:50:30.6 & $18.305 \pm 0.006$ & -- & $16.206 \pm 0.028$ & 4 & M6.5 & Candidates \\ 
611256584470792342 & -- & 03:30:52.74 & $-$27:35:37.6 & $20.290 \pm 0.052$ & $19.098 \pm 0.090$ & $18.449 \pm 0.163$ & 6 & M6.9 & Candidates \\ 
609781520902652982 & -- & 03:30:56.42 & $-$28:28:31.4 & $18.584 \pm 0.007$ & $17.354 \pm 0.027$ & $16.519 \pm 0.037$ & 8 & M6.8 & Candidates \\ 
611256584470790562 & -- & 03:31:10.97 & $-$27:40:47.5 & $22.188 \pm 0.153$ & $19.870 \pm 0.182$ & $17.155 \pm 0.053$ & 4 & L9.2 & Candidates \\ 
611254385447538312 & -- & 03:32:54.89 & $-$28:02:24.0 & $20.906 \pm 0.018$ & $19.109 \pm 0.137$ & $18.063 \pm 0.111$ & 8 & M8.6 & Candidates \\ 
611256447031839364 & -- & 03:32:58.22 & $-$27:35:28.0 & $18.903 \pm 0.012$ & $17.673 \pm 0.025$ & $16.832 \pm 0.042$ & 7 & M6.4 & Candidates \\ 
609788117972418932 & -- & 03:33:14.57 & $-$28:37:58.6 & $21.390 \pm 0.161$ & -- & $16.707 \pm 0.046$ & 4 & L6.0 & Candidates \\ 
609788049252941933 & -- & 03:33:53.91 & $-$28:37:44.0 & $21.751 \pm 0.199$ & -- & $17.192 \pm 0.056$ & 4 & L7.3 & Candidates \\ 
592908071865418214 & EUCL J035236.14-483012.4 & 03:52:36.14 & $-$48:30:12.4 & $20.612 \pm 0.087$ & -- & $18.508 \pm 0.125$ & 2 & L6.0 & Candidates \\ 
591818593281245219 & EUCL J035426.27-491902.2 & 03:54:26.26 & $-$49:19:02.1 & $22.461 \pm 0.255$ & $20.479 \pm 0.267$ & $17.566 \pm 0.055$ & 4 & L8.3 & Candidates \\ 
592915356129954644 & EUCL J035546.73-482350.3 & 03:55:46.73 & $-$48:23:50.3 & $20.950 \pm 0.027$ & $19.860 \pm 0.149$ & $18.660 \pm 0.147$ & 5 & M6.1 & Candidates \\ 
592913706862511090 & EUCL J035918.90-484701.8 & 03:59:18.89 & $-$48:47:01.8 & $21.137 \pm 0.067$ & $19.761 \pm 0.155$ & $18.854 \pm 0.170$ & 6 & M7.1 & Candidates \\ 
614429775028552304 & -- & 06:18:17.19 & $-$25:00:53.7 & $20.300 \pm 0.022$ & $19.168 \pm 0.189$ & $18.332 \pm 0.155$ & 5 & M6.2 & Candidates \\ 
614428331919542481 & -- & 06:18:40.50 & $-$25:15:59.4 & $20.812 \pm 0.070$ & $19.208 \pm 0.192$ & $17.885 \pm 0.103$ & 5 & L2.6 & Candidates \\ 
614430393503843644 & -- & 06:18:55.25 & $-$24:48:33.6 & $19.000 \pm 0.006$ & -- & $16.989 \pm 0.047$ & 4 & M6.7 & Candidates \\ 
614431011979134499 & -- & 06:19:15.69 & $-$24:38:38.9 & $20.079 \pm 0.022$ & $18.836 \pm 0.135$ & $18.044 \pm 0.120$ & 4 & M6.2 & Candidates \\ 
614437059293086812 & -- & 06:19:55.08 & $-$25:00:08.4 & $18.504 \pm 0.003$ & $17.227 \pm 0.032$ & $16.279 \pm 0.029$ & 6 & M8.0 & Candidates \\ 
614438433682623385 & -- & 06:20:20.73 & $-$24:34:40.8 & $19.809 \pm 0.013$ & $18.597 \pm 0.108$ & $17.775 \pm 0.093$ & 5 & M6.5 & Candidates \\ 
612929216534546916 & -- & 06:20:30.98 & $-$25:24:11.9 & $20.192 \pm 0.024$ & -- & $18.166 \pm 0.125$ & 4 & M6.7 & Candidates \\ 
614438364963145565 & -- & 06:20:41.31 & $-$24:37:40.3 & $20.495 \pm 0.023$ & $19.483 \pm 0.242$ & $18.516 \pm 0.173$ & 5 & M6.6 & Candidates \\ 
614437677768379527 & -- & 06:20:53.26 & $-$24:44:30.2 & $19.963 \pm 0.012$ & $18.769 \pm 0.133$ & $17.834 \pm 0.095$ & 5 & M6.3 & Candidates \\ 
612929147815070278 & -- & 06:20:55.25 & $-$25:22:38.3 & $19.779 \pm 0.021$ & -- & $17.165 \pm 0.058$ & 4 & M7.1 & Candidates \\ 
614437609048900375 & -- & 06:21:24.47 & $-$24:51:03.8 & $20.045 \pm 0.013$ & -- & $18.058 \pm 0.107$ & 4 & M6.7 & Candidates \\ 
614436853134657632 & -- & 06:22:03.29 & $-$24:54:52.7 & $19.655 \pm 0.014$ & -- & $17.835 \pm 0.098$ & 4 & M6.6 & Candidates \\ 
614436853134656346 & -- & 06:22:16.54 & $-$24:59:55.8 & $19.277 \pm 0.014$ & -- & $17.186 \pm 0.058$ & 4 & M6.6 & Candidates \\ 
\enddata 
\tablenotetext{a}{Only the photometric values with good-quality flags (Section~\ref{sec:lsst_dp1}) are shown in the table.} 
\tablenotetext{b}{Spectral types of ultracool dwarf candidates are taken from the \texttt{regressor}'s predictions, except for LSST J035236.14$-$483012.4, where we adopt the best-fitting template from its low-S/N Euclid NISP spectrum (Section~\ref{subsec:agreed}). Spectral types of confirmed discoveries are determined from the best-fitting templates to their Euclid spectra.} 
\end{deluxetable*} 
} 
\end{longrotatetable}
}

\end{CJK*}

\clearpage
\bibliographystyle{aasjournal}
\bibliography{ms}{}


\end{document}